\pdfoutput=1
\documentclass[aps,pra,amsmath,amssymb,twocolumn,showpacs,10pt]{revtex4-1}
\usepackage{amsmath}
\usepackage{amssymb}
\usepackage[english]{babel}
\usepackage{graphicx}
\usepackage{dcolumn}
\usepackage{bm}
\usepackage[pdftex]{hyperref}
\usepackage{url}
\usepackage{color}
\usepackage{subcaption}

\begin{document}

\title{Dynamics of one-dimensional quantum droplets}
\author{G. E. Astrakharchik$^{1}$}
\author{B. A. Malomed$^2$}

\begin{abstract}
The structure and dynamics of one-dimensional binary Bose gases forming quantum droplets is studied by solving the corresponding amended Gross-Pitaevskii equation. Two physically different regimes are identified, corresponding to small droplets of an approximately Gaussian shape and large ``puddles'' with a broad flat-top plateau. Small droplets collide quasi-elastically, featuring the soliton-like behavior. On the other hand, large colliding droplets may merge or suffer fragmentation, depending on their relative velocity. The frequency of a breathing excited state of droplets, as predicted by the dynamical variational approximation based on the Gaussian ansatz, is found to be in good agreement with numerical results. Finally, the stability diagram for a single droplet with respect to shape excitations with a given wave number is drawn, being consistent with preservation of the Weber number for large droplets.
\end{abstract}

\date{July 30, 2018}

\affiliation{$^{1}$Departament de F\'{\i}sica, Universitat Polit\`{e}cnica de Catalunya, Barcelona, Spain}
\affiliation{$^2$Department of Physical Electronics, School of Electrical Engineering, Faculty of Engineering and The Center for Light-Matter Interaction, Tel Aviv University, Tel Aviv University, Tel Aviv 69978, Israel}

\pacs{03.75.Kk; 03.75.Mn; 03.75.Lm; 03.75.Hh}

\maketitle

\section{Introduction}

One of the important recent achievements in the studies of ultracold bosonic gases and superfluids is the realization of quantum droplets in a number of experiments with anisotropic interactions between dipolar atoms\cite{Kadau2016,Ferrier-Barbut2016,Chomaz2016,Schmitt2016}, as well as in two-component Bose gases with contact isotropic interactions\cite{LeticiaExperiment1,Semeghini2017arXiv171010890S,LeticiaExperiment2}. Accordingly, the attractive and repulsive forces, whose interplay leads to the formation of quantum droplets, are anisotropic for dipoles and isotropic for the mixture of two-component gases.
In one-dimensional (1D) geometry the (slightly) repulsive mean-field (MF) contribution to the energy per particle scales linearly with density $n$ of the gas, getting balanced by the attractive beyond mean-field (BMF) one, scaling, as $-n^{1/2}$. As a result, the system's energy features a minimum, corresponding to the formation of a liquid droplet\cite{Petrov2015,PetrovAstrakharchik2016}. Notably, its density can be tuned in a wide range, making it possible to create extremely dilute liquids and thus realize, perhaps, the most dilute liquid ever observed in any physical setting. An additional interest in this new class of quantum liquids, as compared to liquid helium, is that the condensate fraction is very large, permitting one to make accurate quantitative predictions based on the mean-field theory amended by the BMF correction. Indeed, the description in terms of the effective Gross-Pitaevskii equation (GPE), used to model the dipolar condensates\cite{WachtlerSantos2016,Bisset2016,WachtlerSantos2016b}, agrees with \textit{ab initio} quantum Monte Carlo calculations for dipolar droplets\cite{Saito2016pimc,Macia2016,Cinti2017}, and for ones formed in the binary BEC dominated by the contact interactions\cite{Cikojevic2017}. It was argued that the quantum droplets may find an application to the design of a precise matter-wave interferometer\cite{McDonald2014,Laburthe-Tolra2016,Lepoutre2016}.

While the present study concentrates on the properties in one-dimensional geometry, we find it instructive to make a comparison with a three-dimensional (3D) counterpart in terms of the sign of the BMF corrections and the value of the gas parameter where MF theory can be applied. Indeed, one-dimensional systems might seem counterintuitive in certain aspects. Suppose we consider a single-component gas with delta-interacting potential $V(r) = g \delta (r)$, its potential energy per particle is $E/N = g n g_2(0)/2$ where $g_2(0)$ is the value of the density-density correlation function $g_2(r) =\langle n(r) n(0)\rangle / \langle n\rangle^2$ at contact position $r=0$. The potential energy per particle scales linearly with the density. The potential energy can be reduced to zero by making the particle fully impenetrable, $g_2(0) = 0$. On the other hand in one dimension the impenetrable condition induces kinetic energy per particle which scales quadratically with the density, $E/N \propto \hbar^2 n^2 / m$. It means that in one dimension the mean field regime, where one can neglect correlations and set $g_2(0) = 1$, is reached for large density. Here the mean-field energy $n$ becomes smaller than $n^2$ dependence of a strongly correlated (Tonks-Girardeau) gas which is obtained when $g_2(0)=0$ (Pauli exclusion). This is exactly opposite to the ``usual'' three dimensional situation where the mean-field energy scaling $\propto n$ becomes energetically preferable at small densities, compared to the kinetic energy per particle due to Pauli principle $\propto \hbar^2 n^{2/3}/m$. As a result, the regimes of the applicability of the mean-field theory are swapped and correspond to small (3D) and large (1D) densities.

Another important difference is the sign and the structure of the beyond-mean field terms. In three dimensions the BMF correction was first calculated for a single-component 3D hard sphere gas by Lee-Huang-Yang\cite{Huang57,Lee57} back in 1957. The textbook derivation\cite{LifshitzPitaevskiiBook} is based on the Bogoliubov theory. In 1D, the energy of the $\delta$-pseudopotential gas was obtained by Lieb and Liniger\cite{LiebLiniger63} in 1963 for arbitrary interaction strength by using Bethe ansatz for the ground-state wave function and by relating the energy to an integral over solution of Love integral equations\cite{LoveEquation1949}. A perturbative solution to such integral equations has shown that the Bogoliubov theory reproduces correctly the leading MF and subleading BMF terms. Such verification is important as it justifies the use of MF and BMF theory while strictly speaking, the condensate fraction is zero in 1D due to the Hohenberg theorem\cite{Hohenberg67}.

Although the BMF terms in 1D are sometimes loosely referred to as LHY terms, actually neither of Lee, Huang or Yang ever calculated them. It is rather curious to note that the related expressions were obtained by Kirchhoff in 1877 in a different problem\cite{Kirchhoff1877}. He was calculating the capacitance of a circular capacitor as a function of the distance between two circular plates. The capacitance can be calculated to an integral over the solution of Love integral equations\cite{LoveEquation1949}, that is exactly the same equations as those appearing in the Bethe ansatz theory. Indeed, within the electrostatic analogy the density $n$ of 1D Bose gas can be mapped to the conductance of the circular plates, the energy $E$ to the second moment, and the coupling constant $g$ is proportional to the radius of each plate\cite{Gaudin71}. Kirchhoff comments that ``Their computation is generally cumbersome because it requires finding solution of entangled, transcendental equations; but it is very simple if one'' takes the limit of small separation.
In the leading term, the capacitor charge is inversely proportional to the separation, resulting in a linear dependence of the energy per particle on the density. That is, the standard mean-field Gross-Pitaevskii result is recovered with the energy per particle proportional to the density, $E/N = gn/2$. In the subleading order, the edges of the capacitor plates have to be considered. With the same accuracy as Kirchhoff uses the energy 
per particle becomes\cite{Gaudin71} $E/N = gn/2 - 2/(3\pi)\;g^{3/2}\sqrt{n m}/\hbar$.
That is, the subleading term is negative and proportional to $\sqrt{n}$.
This correction exactly coincides with beyond-mean-field terms as found within the Bogoliubov theory\cite{LiebLiniger63} and as a 1D analog of LHY terms.

The same conclusion about the sign can be obtained by a ``hand waving'' argument by noting that MF scaling $propto g_2(0) n, n\to\infty$ can be smoothly matched to $\propto n^2, n\to 0$ Tonks-Girardeau energy if the BMF correction is negative. This corresponds to decreasing gradually $g_2(0)$ from 1 (MF) to 0 (TG). A more rigorous explanation for the sign is that the BMF energy is obtained in a second-order perturbation theory which has to reduce the energy. The situation is somewhat different in 3D as within the second-order theory the relation between the coupling constant $g$ and the $s$-wave scattering length must be corrected in comparison with the simple MF expression eventually resulting in a positive LHY correction in three dimensions\cite{Huang57,Lee57}.

The equilibrium densities of both components of the binary condensate depend on interaction strengths and atomic masses of the two species. The symmetric case of equal masses and strengths of intraspecies interactions, with equal numbers of atoms in each components, allows a simpler and more elegant description. In this case, the density profiles of both components coincide and can be described by the effective three-dimensional (3D) single-component GPE with cubic and quartic nonlinearities, the latter term representing the LHY correction\cite{Petrov2015}. This possibility provides an important interdisciplinary connection to the field of nonlinear optics\cite{KivsharBorisRevModPhys1989}, as concerns the model equations with higher-order nonlinearities\cite{AstrakharchikAE1995,Cid1,Cid2} and, possibility, controlled generation of solitons in these systems. On the other hand, in the case when two-component features in the dynamics are essential, they may be affected by an additional linear interconversion between the components\cite{Cappellaro2017}.

In three and two dimensions, quasi-1D solitons are unstable with respect to the transverse snake instability, although the stability can be enhanced by imposing rotation to the quantum droplets\cite{Boris2017Vortexlattices, Boris2018vortex}. The advantage of the proper 1D geometry, imposed by the tight confinement in the transverse directions (cf. the experimental realization of the Tonks-Girardeau gas\cite{Paredes04,Kinoshita04}), is that such an instability is absent, thus permitting one to realize a very clean and highly controllable many-body test bed which may permit the measurement of quantum many-body effects with very high precision.

Commonly known hallmarks of solitons are being (i) self-trapped and (ii) robust with respect to soliton-soliton collisions. While the former feature is definitely present in quantum droplets, the latter one should be yet verified. It was proposed to use Gaussian \textit{ans\"{a}tze} for gaining an analytical insight in physics of dipolar\cite{WachtlerSantos2016b} and BEC\cite{LeticiaExperiment1} droplets. In particular, the dynamical version of the Gaussian-based variational approximation (VA)  can be used to predict the frequency of intrinsic oscillations of the soliton-like objects in an excited state\cite{Anderson,Salasnic2000,Progress}. Excitations in a dipolar quantum droplet have been experimentally studied and a scissors mode has been observed in it\cite{Ferrier-Barbut2017arXiv171206927F}. Recently, intrinsic modes were theoretically investigated in Fermi-Bose mixtures\cite{Karpiuk2018arXiv180100346K}, spin-orbit-coupled Bose-Einstein condensates (BECs) \cite{Boris2017spinorbit}, including those dominated by the LHY terms\cite{NJP}, and in a discrete BEC model  \cite{Boris2017doublewell}.

The collisions between dipolar droplets were experimentally studied in Ref.~\cite{Pfau2016JPB}.
The system was confined to an elongated trap and the interactions were quenched in such a way that the density distribution was split into multiple pieces.
The droplets which have been formed were shown to be long lived and their dynamics was studied.
Until now no similar experiment has been performed with short-range interacting droplets although such experimental studies might be expected in the near future.
The goal of our study is to analyze the dynamic properties of ultradilute quantum droplets.

In the present work we address collisions of 1D quantum droplets and intrinsic oscillations of an isolated droplet, along with excitations generated by imprinting onto it a density modulation with a certain wave number. The article is organized as follows. First, we address static droplets in Sec.~\ref{Sec:static properties}, where we start by introducing the model in subsection~\ref{Sec:model}. Subsection~\ref{Sec:limits} addresses the asymptotic analysis in the limits of small and large droplets. In subsection~\ref{Sec:var} we develop the Gaussian \textit{ansatz} for the study of both stationary and dynamical properties of the droplet. The conditions of applicability of the Gross-Pitaevskii equation for describing statical and dynamical properties are analyzed in Sec.~\ref{Sec:conditions of applicability}. We consider energetic and spatial properties of stationary states in subsection~\ref{Sec:energy}, minimal number of atoms necessary for the creation of the droplet in subsection~\ref{Sec:minimal number}, and calculate the surface tension in Subsection~\ref{Sec:surface tension}. Section~\ref{Sec:time dynamics} addresses dynamical effects. In Subsection~\ref{Sec:collisions} we consider collisions of two droplets, and the dynamics of a single one is considered in subsection~\ref{Sec:single droplet dynamics}, including intrinsic monopole oscillations in subsubsection~\ref{Sec:breathing mode}, and excitations generated by periodic density modulation in subsubsection~\ref{Sec:finite momentum}. The stability diagram for the quantum diagram is produced in subsection~\ref{Sec:stability diagram}. We finish by drawing conclusions in Sec.~\ref{Sec:conclusions}.

\section{Static quantum droplets\label{Sec:static properties}}

In this Section we address static equilibrium properties of a single droplet by considering the exact solution to the BMF-amended GPE, as well as getting an additional insight from the Gaussian-based variational approximation (VA).

\subsection{Model system\label{Sec:model}}

We consider the binary BEC with mutually symmetric spinor components, assuming that the coupling constants describing the repulsion between the atoms in each one are equal, $g_{\uparrow\uparrow}=g_{\downarrow\downarrow}\equiv g$, and numbers of atoms in the components are equal too. In this case, the equilibrium densities of both components are identical, which makes the analysis essentially easier, and results clearer.

The underlying time-dependent GPE for the one-dimensional droplet with the symmetric components is~\cite{PetrovAstrakharchik2016}
\begin{equation}
i\hbar\psi_t=-\frac{\hbar^2}{2m}\psi_{xx}+\delta g|\psi |^2\psi -\frac{\sqrt{2m}}{\pi \hbar}g^{3/2}|\psi |\psi \;,
\label{Eq:GPE:full units}
\end{equation}
where parameters $\delta g$ and $g$ are positive and are related to the coupling constants in the two spinor components as $\delta g=g_{\uparrow \downarrow}+\sqrt{g_{\uparrow \uparrow}g_{\downarrow \downarrow}}>0$ and $g=\sqrt{g_{\uparrow \uparrow}g_{\downarrow \downarrow}}$. The coupling constant $g$ is relevant for inducing a hard ``spin'' mode while the difference $\delta g$ between attractive intercomponent and repulsive intracomponent interactions is responsible for appearance of a soft ``density'' mode and condition $\delta g\ll g$ induces a separation of scales.

In experiments it is possible to tune $\delta g$ both to positive or negative values. The proper sign is chosen in such a way that the imbalance in the mean-field terms is opposite to the beyond mean-field contribution and, consequently, the imbalance depends on dimensionality of the problem. In one dimension, the beyond mean-field terms are directly obtained from the second-order perturbation theory which produces a negative correction to the energy\cite{LiebLiniger63}. Accordingly, a positive mean-field imbalance is needed, $\delta g>0$, for producing am energy minimum in the equation of state. In 3D, the BMF term includes the renormalization correction\cite{LifshitzPitaevskiiBook} to the scattering amplitude within the second Born approximation, resulting in the positive LHY term and requiring $\delta g<0$\cite{Petrov2015}.

We define characteristic units of length $x_{0}$, time $t_{0}$ and energy $E_{0}$:
\begin{eqnarray}
x_{0} &=&\frac{\pi \hbar^2\sqrt{\delta g}}{\sqrt{2}mg^{3/2}},
\label{Eq:units:x0} \\
t_0 &=&\frac{\pi^2\hbar^{3}\delta g}{2mg^{3}}, \\
E_0 &=&\frac{\hbar^2}{mx_0^2}=\frac{\hbar}{t_0}=\frac{2mg^{3}}{\pi^2\hbar^2\delta g}\;,
\label{Eq:units}
\end{eqnarray}
which yield a characteristic factor for the normalization of the wave function,
\begin{equation}
\psi_{0}=\frac{\sqrt{2m}}{\pi \hbar \delta g}g^{3/2}\;.
\label{Eq:psi0}
\end{equation}
We demonstrate below that
\begin{equation}
N_{0} = \psi_{0}^2 x_0 = \frac{\sqrt{2}}{\pi}\left(\frac{g}{\delta g}\right)^{3/2}\;.
\label{Eq:No}
\end{equation}
determines a critical number of particles separating two different physical regimes.

Thus, rescaling
\begin{equation}
t=t_{0}t^{\prime},~x=x_{0}x^{\prime},~\psi =\psi_0 \psi^{\prime}
\label{Eq:scaling}
\end{equation}
casts Eq.~(\ref{Eq:GPE:full units}) in an equation without free coefficients (where the primes are omitted):
\begin{equation}
i\psi_{t}+\frac{1}{2}\psi_{xx}-|\psi |^2\psi +|\psi |\psi =0,
\label{Eq:GPE}
\end{equation}

A peculiarity of the 1D geometry is that the ground-state solution of the GPE for the droplet, Eq.~(\ref{Eq:GPE}), can be found in an explicit form\cite{PetrovAstrakharchik2016}:
\begin{equation}
\psi_{\mathrm{exact}}(x)=-\frac{3\mu \exp \left( -i\mu t\right)}{1+\sqrt{1+\frac{9\mu}{2}}\cosh (\sqrt{-2\mu x^2})}\;,
\label{exact:psi}
\end{equation}
with the relation between normalization $N$ and chemical potential $\mu $ given by
\begin{equation}
N=\frac{4}{3}\left[ \ln \left( \frac{\sqrt{-\frac{9}{2}\mu}+1}{\sqrt{\frac{9}{2}\mu +1}}\right) -\sqrt{-\frac{9}{2}\mu}\right] .  \label{Eq:mu}
\end{equation}
The equilibrium density corresponding to the spatially uniform state ($N\rightarrow \infty $), and the respective chemical potential, in units defined by Eqs.~(\ref{Eq:units:x0}) and (\ref{Eq:units}), are
\begin{equation}
\frac{n_{0}}{\psi_{0}^2}=\frac{4}{9}\;,  \label{exact:n0}
\end{equation}
\begin{equation}
\frac{\mu_{0}}{E_{0}}=-\frac{2}{9}\;.  \label{Eq:mu:bulk}
\end{equation}

\subsection{Limit cases of small and large droplets\label{Sec:limits}}

In a large finite-size droplet (``puddle''), $\mu$ approaches the constant value~(\ref{Eq:mu:bulk}) corresponding to the chemical potential of an infinitely extended uniform liquid at zero pressure.
The chemical potential~(\ref{Eq:mu}) is expanded as
\begin{equation}
\mu =-\frac{2}{9}+\frac{8}{9}\exp \left(-2-\frac{3}{2}N\right) ,
\label{exact:mu:largeN}
\end{equation}
and features an exponentially weak dependence on $N$. On the other hand, for small droplets with small $N$ the dependence has a power-law form:
\begin{equation}
\mu =-\frac{1}{2^{1/3}3^{2/3}}N^{2/3}=-0.382N^{2/3}\;.
\label{exact:mu:smallN}
\end{equation}
In this case, the dependence on $N$ is much stronger, as long as $|\mu|$ is small.

The total energy $E$ can be obtained by integrating the chemical potential, $E(N)=\int_{0}^{N}\mu (N^{\prime})dN^{\prime}$.
For small $N$, Eq.~(\ref{exact:mu:smallN}) results in a power-law dependence,
\begin{equation}
E=-\frac{1}{5}\left( \frac{3}{2}\right)^{1/3}N^{5/3}=-0.229N^{5/3}\;,
\label{exact:E:smallN}
\end{equation}
while Eq.~(\ref{exact:mu:largeN}) produces an asymptotically linear dependence on large $N$:
\begin{equation}
E=-\frac{6}{27}N+\frac{16\exp (-2)}{27}-\frac{16}{27}\exp \left( -2-\frac{3}{2}N\right) \;.  \label{exact:E:largeN}
\end{equation}

A typical size of the droplet can be easily estimated in both limits. The large droplet includes a bulk (\textit{flat-top}) region with the nearly uniform density given by Eq.~(\ref{exact:n0}), with size $L=N/n_{0}$. The respective mean-square size also increases linearly with the number of particles,
\begin{equation}
\sqrt{\langle x^2\rangle}=\frac{L}{2\sqrt{3}}=\frac{N}{2\sqrt{3}n_{0}}=0.65\;.  \label{x2:largeN}
\end{equation}

\subsection{The variational approximation (VA) based on the Gaussian \textit{ansatz}\label{Sec:var}}

In this subsection we approximate the shape of the droplet by a Gaussian and optimize its width according to the variational principle. This simple model provides additional insight in properties of the droplets as the number of particles varies.

\begin{figure}[tbp]
\includegraphics[width=\columnwidth,angle=0]{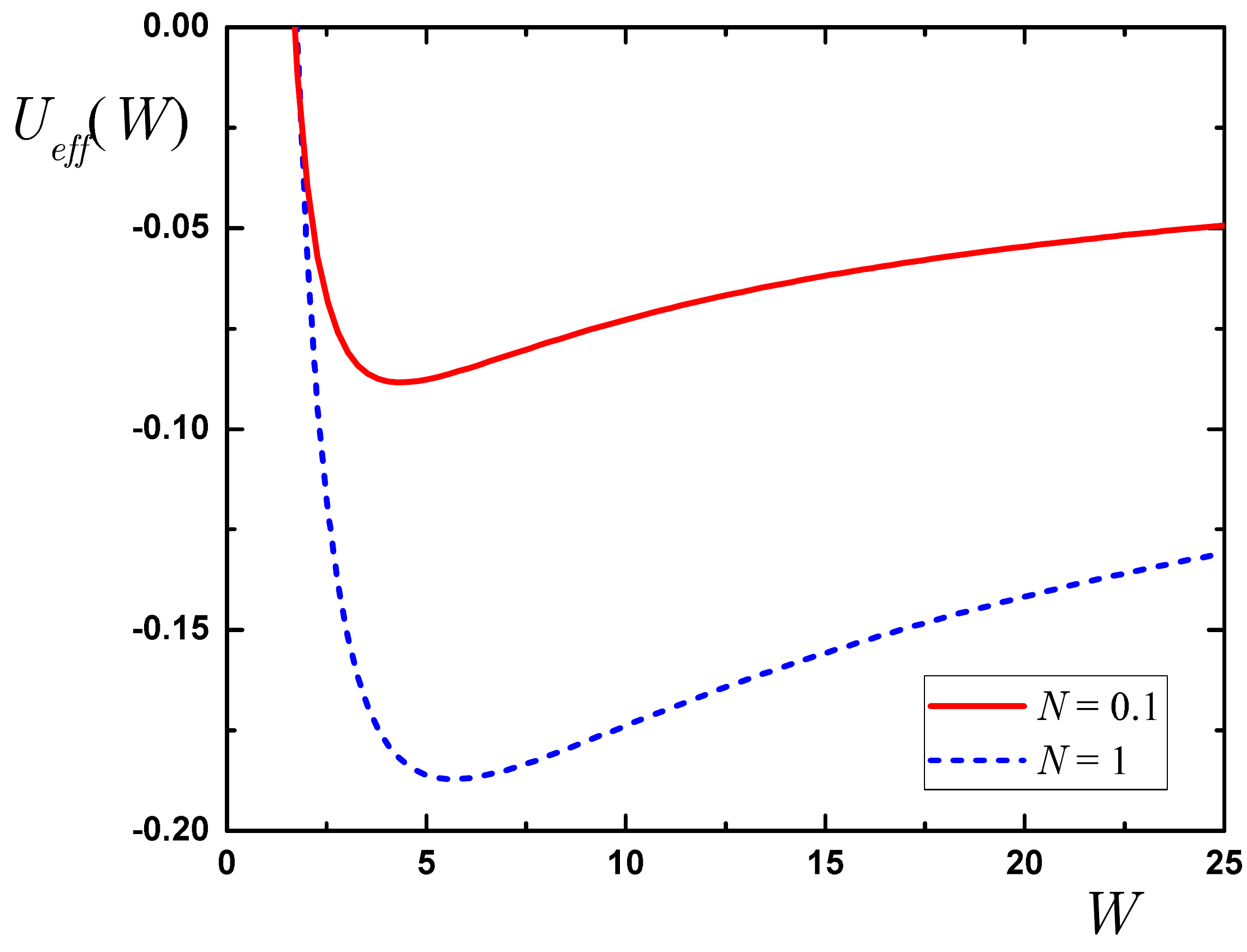}
\caption{(Color online) The plot of the effective potential~(\ref{Eeff}) for $N=0.1$ and $N=1$. The minimum of the potential defines the optimal width $W$ of the droplet.}
\label{Fig:Ueff}
\end{figure}

The VA is based on the Lagrangian for Eq.~(\ref{Eq:GPE}),
\begin{gather}
L=\int_{-\infty}^{+\infty}\mathcal{L}dx,  \label{L} \\
\mathcal{L}=\frac{i}{2}\left( \psi \psi_{t}^{\ast}-\psi^{\ast}\psi
_{t}\right) +\frac{1}{2}\left\vert \psi_{x}\right\vert^2+\frac{1}{2}
\left\vert \psi \right\vert^{4}-\frac{2}{3}|\psi |^{3}.  \label{density}
\end{gather}
To develop the dynamical version of the VA, we adopt the Gaussian \textit{ansatz},
\begin{equation}
\psi =A(t)\exp \left[ i\phi (t)-\frac{x^2}{2\left( W(t)\right)^2}+ib(t)x^2\right] \;,  \label{Eq:Gaussian ansatz}
\end{equation}
where $A$, $\phi $, $W$, and $b$ are real amplitude, phase, width and chirp, respectively (in the time-independent version of the VA, $b=0$). Although the large-distance Gaussian asymptotic form of wave function~(\ref{Eq:Gaussian ansatz}) is incompatible with the exponential decay of the exact solution~(\ref{exact:psi}), we demonstrate below that the overall accuracy provided by the VA is extremely good. The normalization of the wave function determines the number of particles in the droplet,
\begin{equation}
N=\int_{-\infty}^{+\infty}\left\vert \psi (x)\right\vert^2dx=\sqrt{\pi}A^2W\;.  \label{N}
\end{equation}
Substituting the \textit{ansatz} in Lagrangian density~(\ref{density}) and using Eq.~(\ref{N}) to eliminate $A^2$ in favor of $N$, as $A^2=N/\sqrt{\pi}W$, one can produce the effective Lagrangian:
\begin{equation}
\frac{L_{\mathrm{VA}}}{N}=\frac{d\phi}{dt}+\frac{W^2}{2}\frac{db}{dt}+\frac{1}{4W^2}+W^2b^2+\frac{N}{2\sqrt{2\pi}W}-\frac{2^{3/2}\sqrt{N}}{3^{3/2}\pi^{1/4}\sqrt{W}}\;.
\end{equation}
The Euler-Lagrange equations for variables $b$ and $W$ are derived from
here:
\begin{equation}
b=\frac{1}{2W}\frac{dW}{dt}\;,  \label{b}
\end{equation}
\begin{equation}
\frac{d^2W}{dt^2}=\frac{1}{W^{3}}+\sqrt{\frac{1}{2\pi}}\frac{N}{W^2}-\frac{2^{3/2}N^{1/2}}{\pi^{1/4}(3W)^{3/2}}\equiv -\frac{dU_{\mathrm{eff}}}{dW}\;,  \label{Eq:W}
\end{equation}
where the effective potential for oscillations of the soliton's width is
\begin{equation}
U_{\mathrm{eff}}(W)=\frac{1}{2W^2}+\sqrt{\frac{1}{2\pi}}\frac{N}{W}-\frac{2}{\pi^{1/4}}\left( \frac{2}{3}\right)^{3/2}\sqrt{\frac{N}{W}}.
\label{Eeff}
\end{equation}
This potential gives rise to a shallow well, as shown in Fig.~\ref{Fig:Ueff}

In the framework of the VA, the stationary soliton corresponds to the minimum of potential~(\ref{Eeff}). The soliton's width is determined by the cubic equation for $\sqrt{W}$, which follows from condition for the potential minimum, $dU_{\mathrm{eff}}/dW=0$, as per Eq.~(\ref{Eq:W}):
\begin{equation}
1+\sqrt{\frac{1}{2\pi}}{NW}-\frac{1}{\pi^{1/4}}\left( \frac{2}{3}\right) ^{3/2}\sqrt{N}W^{3/2}=0\;.  \label{equil}
\end{equation}
In particular, in both asymptotic limits of $N\rightarrow 0$ and $N\rightarrow \infty $ the width is large:
\begin{eqnarray}
W(N &\rightarrow &0)\approx \frac{3}{2}\pi^{1/6}N^{-1/3}\;,
\label{Eq:width:Nsmall} \\
W(N &\rightarrow &\infty )\approx \frac{27}{16\sqrt{\pi}}N\;.
\label{Eq:width:Nlarge}
\end{eqnarray}
It is also possible to find the exact \emph{minimum value} of the VA-predicted width,
\begin{equation}
W_{\min}=\frac{3\sqrt{3}}{2^{3/4}}\approx 3.09 \label{Eq:width:Wmin}
\end{equation}
which is attained at the value of the norm of the order of one, namely
\begin{equation}
N\left(W_{\min}\right) = \frac{2^{5/4}\sqrt{\pi}}{3^{3/2}}\approx 0.81\;. \label{N-min}
\end{equation}

The mean-square size of the droplet with the Gaussian profile is $\langle x^2\rangle =W^2(N)/2$. In the $N\rightarrow 0$ limit, the width is given by Eq.~(\ref{Eq:width:Nsmall}), the respective mean-square size increasing as $N^{-1/3}$:
\begin{equation}
\sqrt{\langle x^2\rangle}=\frac{3}{2\sqrt{2}}\pi^{1/6}N^{-1/3}\approx 1.28~N^{-1/3}\;.\label{x2:smallN}
\end{equation}
Comparison with Eq.~(\ref{x2:largeN}) makes it evident that the droplet has a minimum size at $N\sim 1$.

The frequencies of low-lying collective excitations are determined by the eigenvalues of the Hessian matrix evaluated at the equilibrium position\cite{Salasnic2000}. In the 1D system, it reduces to the second derivative
\begin{equation}
\frac{d^2U_{\mathrm{eff}}(W_{\min})}{dx^2}=\frac{1}{2}\omega^2\;,
\label{Eq:freq}
\end{equation}
and corresponds to the ``breathing'' (compression-dilatation) mode of frequency $\omega$. For limit values of $N$ the frequency takes the approximate form,
\begin{eqnarray}
\omega (N\rightarrow 0) &=&\frac{2\sqrt{\frac{2}{3}}}{3\sqrt[3]{\pi}}
N^{2/3},  \label{Eq:freq:small} \\
\omega (N\rightarrow \infty ) &=&\frac{32~2^{1/4}\sqrt{\frac{\pi}{3}}}{81} \frac{1}{N}\;.  \label{Eq:freq:large}
\end{eqnarray}

The ground-state energy is obtained by evaluating the energy functional
\begin{equation}
E=\int_{-\infty}^{+\infty}\left( \frac{1}{2}\left\vert \psi_{x}\right\vert^2+\frac{1}{2}\left\vert \psi \right\vert^{4}-\frac{2}{3}
|\psi |^{3}\right) dx,
\end{equation}
and the chemical potential is then obtained as its derivative with respect to the number of particles, $\mu =dE/dN$. The chemical potential of a droplet is negative and approaches zero, in the limit of $N\rightarrow 0$, as
\begin{equation}
\mu =-\frac{5N^{2/3}}{9\pi^{1/3}}=-0.379N^{2/3}\;.  \label{Eq:VA:mu}
\end{equation}
Note that the difference of this VA-produced result with the exact one, given by Eq.~(\ref{exact:mu:smallN}), is less than $1\%$.

\subsection{Conditions of applicability\label{Sec:conditions of applicability}}

An important issue is to clarify the regions of applicability of the GPE~(\ref{Eq:GPE:full units}) for describing static and dynamic properties of the droplets. The GPE has been proven to be immensely useful for description of experiments with single-component ultracold Bose gases, as the mean-field description it provides is sufficient for interpretation of a large variety of quantum effects\cite{PitaevskiiStringariBook2016}. On the other hand, artificial incorporation of terms with higher powers of the condensate wave function in the GPE in order to ``simulate'' LHY terms in the energy may lead to incorrect dynamics of excitations in the framework of the amended equation.

The key point here is that the dominant contribution to the BMF energy comes from distances smaller than or comparable to the healing length $\xi$, defined, in terms of the sound velocity, $c$, by $\hbar^2/(2m\xi^2)=mc^2$. In order to treat the BMF term as a ``local'' contribution, the distances at which the density profile changes should be large in comparison to the healing length. In a single-component gas this is impossible, as the density profile changes exactly at distances $\sim \xi$. In quantum droplets, the situation is completely different, as the variation in the density profile is governed by the ``soft''  healing length
\begin{equation*}
\xi_{-}=\frac{\hbar}{\sqrt{2}mc_{-}},
\end{equation*}
while the BMF energy is earned at the distances comparable to the
``hard'' healing length,
\begin{equation*}
\xi_{+}=\frac{\hbar}{\sqrt{2}mc_{+}}\;.
\end{equation*}
The speed of sound of the two modes is defined by $mc_{\pm}^2=(g\pm |g_{\uparrow \downarrow}|)n$ in the symmetric case\cite{Petrov2015,PetrovAstrakharchik2016}. Thus, the coupling constant $\delta g$ defines the speed in the soft mode, $mc_{-}^2=\left(\delta g\right) n$, while $g\gg \delta g$ produces a much larger speed in the hard mode, $mc_{+}^2\approx gn$.

The large separation of scales, $\xi_{-}\gg \xi_{+}$, justifies the inclusion of higher-order terms in the GPE as local ones. Accordingly, the relatively slow dynamics, which takes place at timescales large compared to the typical ``hard'' time interval,
\begin{equation*}
t_{+}=\frac{2m\xi_{+}^2}{\hbar},
\end{equation*}
is correctly described by the amended GPE~(\ref{Eq:GPE:full units}).

\subsection{Static energetic and spatial properties \label{Sec:energy}}

In this subsection we address different physical states that can be realized in a single stationary droplet. Exact solution~(\ref{exact:psi}) provides the necessary information for this. The exact solution can be also used to verify the accuracy of the VA.

\begin{figure}[tbp]
\caption{(Color online) Density profiles for different values of $N$. Exact analytical solution~(\ref{exact:psi}) and predictions produced by Gaussian \textit{ansatz}~(\ref{Eq:Gaussian ansatz}) are shown by solid and dashed lines, respectively. For the smallest value, $N=0.1$, the lines are indistinguishable. For the largest values of $N$, the bulk region, filled by the density with the value given by Eq.~(\ref{exact:n0}), is observed at the center. For these values ($N=10$ and $20$), the VA prediction are not displayed, as they are irrelevant in that case.}
\includegraphics[width=\columnwidth,angle=0]{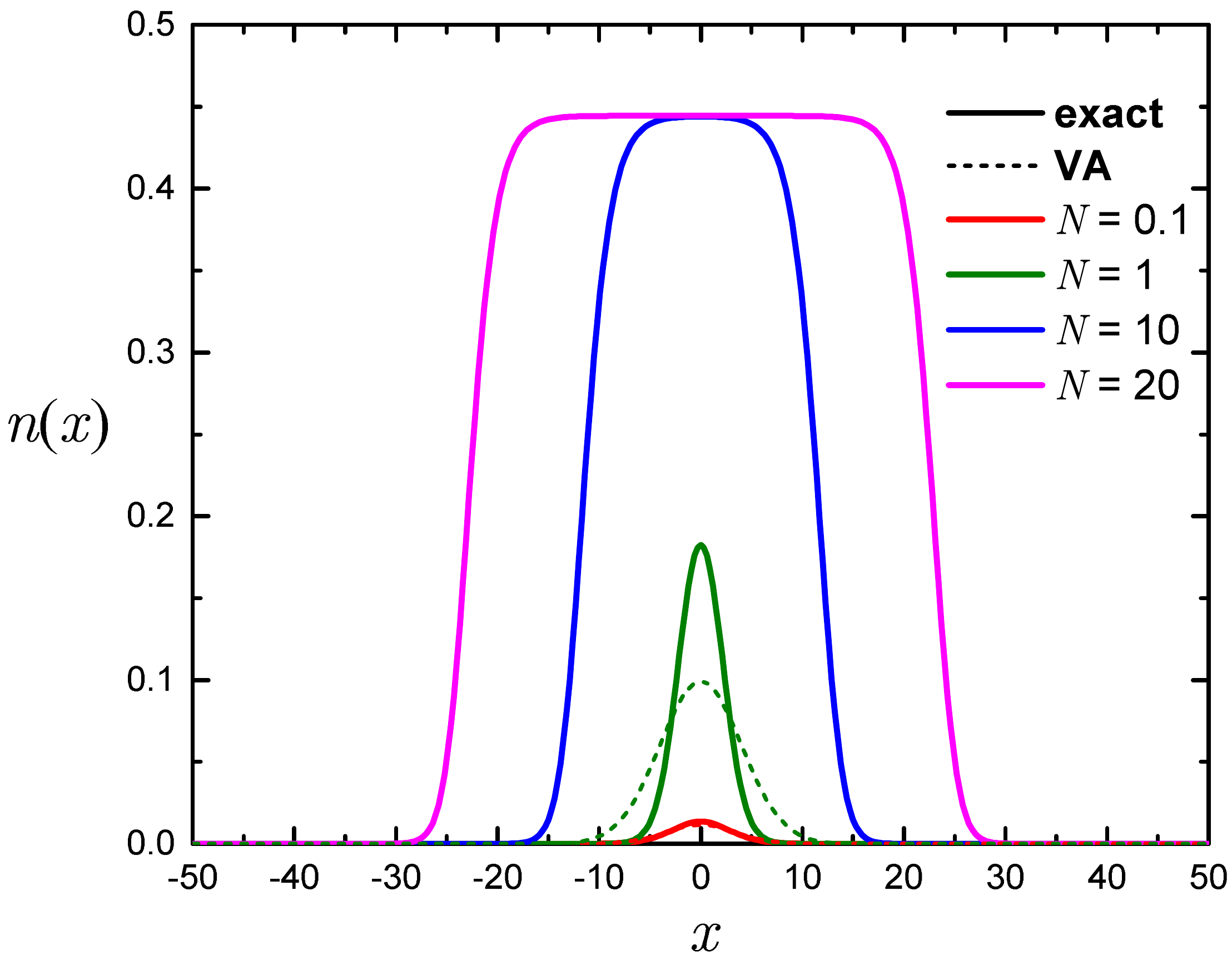}
\label{Fig:density}
\end{figure}

We start by considering the spatial profile of the droplet. A number of characteristic density profiles of static droplets are displayed in Fig.~(\ref{Fig:density}). In the framework of the mean-field theory, profile $n(x)$ is governed by a single parameter, \textit{viz}., dimensionless norm $N$, which is the number of particles divided by $N_0$, see Eq.~(\ref{Eq:No}). For $N\ll 1$, the shape of the droplet is essentially non-uniform, the kinetic term (second derivative) in Eq.~(\ref{Eq:GPE:full units}) being relevant for determining the shape. To a certain extent, this case is similar to that of ``standard'' single-component bright solitons with the cubic nonlinearity, where the quantum pressure balances the potential energy. As $N$ increases, the density at the center of the soliton grows monotonously until it attains the equilibrium bulk value $n_{0} $, see Eq.~(\ref{exact:n0}), at $N\approx 10$. By increasing $N$ further, formation of a ``puddle'' with a flat plateau in its center is observed, which is filled by the bulk phase with density $n_{0}$. The action of the kinetic term is essential near edges of the droplet, being irrelevant in the plateau region. The situation reminds that of a classical liquid, where a plateau in the density profile expands with the growth of the droplet's mass.

The peculiarity of the present model is that it admits two distinct regimes, separated by value $N\sim 1$, i.e., the number of particles $\sim N_0$, see Eq.~(\ref{Eq:No}). For smaller $N$, the quantum pressure is significant, and the density profile is essentially non-uniform, while for larger $N$ the above-mentioned bulk region appears, making the pattern similar to a ``puddle'' filled by the homogeneous liquid.

In addition, we have tested the density profile derived from the Gaussian-based VA. Its prediction is shown in Fig.~\ref{Fig:density} by the dashed line. For $N=0.1$, the Gaussian density profile is almost indistinguishable from the exact solution, which justifies the use of the VA for small droplets.

\begin{figure}[tbp]
\includegraphics[width=\columnwidth,angle=0]{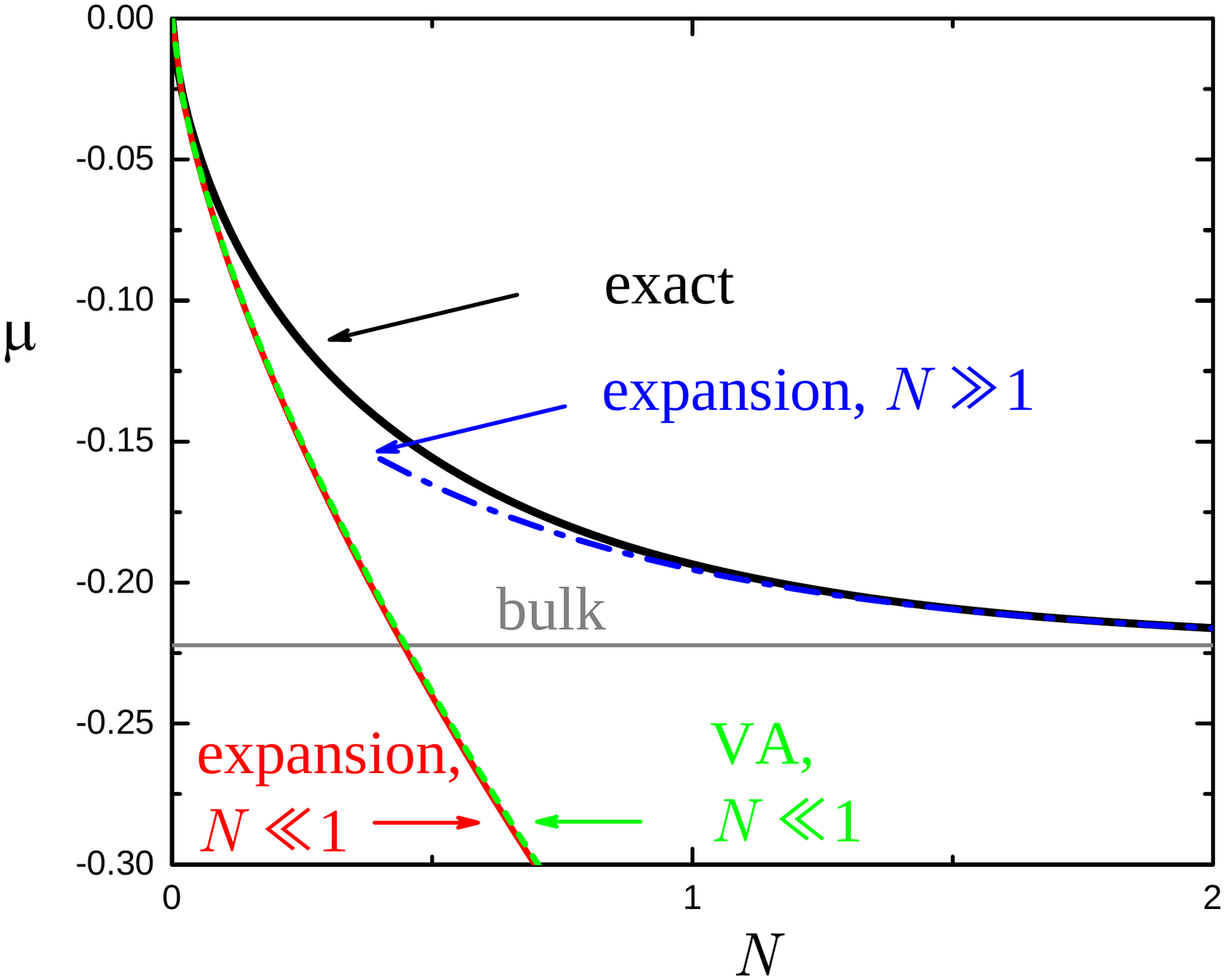}
\caption{(Color online) Chemical potential $\mu$ as a function of norm $N$. The bulk value (\ref{Eq:mu}) of chemical potential is shown by the horizontal line. The large-$N$ expansion, given by Eq.~(\ref{exact:mu:largeN}), is shown by the dashed-dotted line. The small-$N$ expansion, given by Eq.~(\ref{exact:mu:smallN}), and VA prediction~(\ref{Eq:VA:mu}) are hardly distinguishable.}
\label{Fig:mu}
\end{figure}

To address the energetic properties, we report the value of the chemical potential in Fig~\ref{Fig:mu} as a function of norm $N$. The chemical potential is always negative, implying that the state is self-bound in the equilibrium. One can compare the expansions derived for small and large $N$ from both the exact and VA solutions. As anticipated, for small droplets the chemical potential is very well captured by the VA, being close to values given by expansion (\ref{exact:mu:smallN}), which was obtained from the exact solution. As the droplet's size increases, $\mu $ attains the bulk value (\ref{Eq:mu:bulk}) exponentially fast, in agreement with prediction~(\ref{exact:mu:largeN}).

\begin{figure}[tbp]
\caption{(Color online) The mean-square size $\sqrt{\langle x^2\rangle}$ of the droplet vs. norm $N$, compared to the asymptotic expressions given by Eqs.~(\ref{x2:largeN}) and (\ref{x2:smallN}).}
\label{Fig:x2mean}\includegraphics[width=\columnwidth,angle=0]{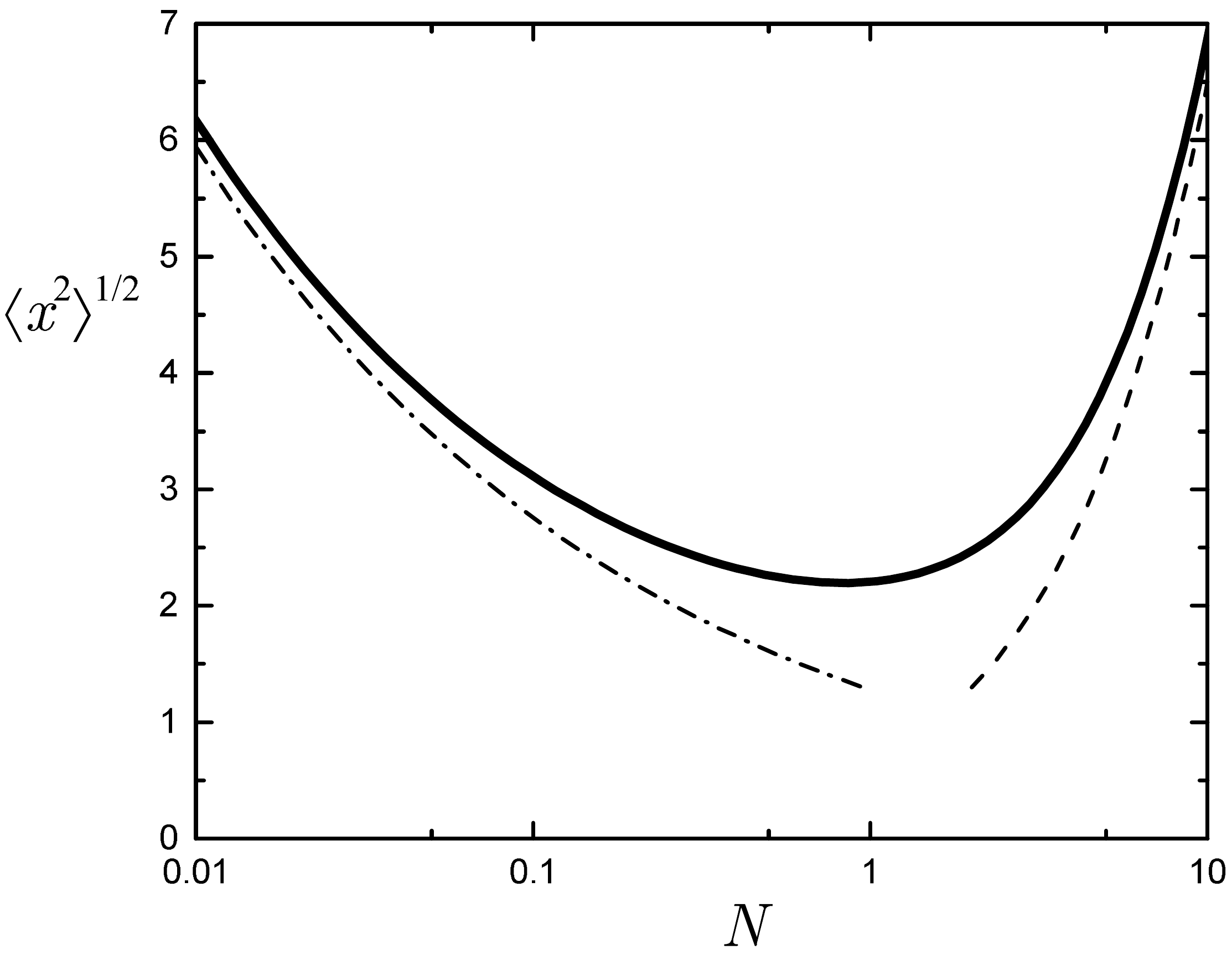}
\end{figure}

We conclude the study of static properties of the droplets by the consideration of their size as a function of the norm. The mean-square size $\sqrt{\langle x^2\rangle}$ of the droplet is reported in Fig.~\ref{Fig:x2mean}. Measured in units of $x_{0}$ [see Eq.~(\ref{Eq:units:x0})], the droplet has a minimal size for $N\approx 1$, i.e., for the number of particles close to $N_0$. A smaller droplet features an approximately Gaussian shape with width $W$ given by Eq.~(\ref{Eq:width:Wmin}), which diverges at $N\rightarrow 0$. In the opposite limit, the size of large droplets grows linearly with $N$. It is worthy to note that, while the VA cannot predict the flat-top shape of the ``puddle'' for $N\gg 1$, it is still able to correctly predict that the droplet's size increases in this regime. The location of the minimum of the width at $N\simeq 1$ implies that $N_0$ [see Eq.~(\ref{Eq:No})] determines the density at which the droplet has the most compact form.

\subsection{The minimal number of atoms in a droplet\label{Sec:minimal number}}

In dimensions lower than three, a purely attractive potential between two particles always leads to formation of a bound state, which is different from the 3D case, where a finite threshold for the formation of a two-body bound state exists. Therefore, the 3D droplets are formed only if the number of atoms exceeds a certain critical number. For $^3$He atoms, a liquid is formed for $N\gtrsim 20$ atoms\cite{Stringari1985}, while in ultradilute quantum droplets the necessary number may be larger\cite{Cikojevic2017}. Contrarily, 1D dimers are always formed for any value of the $s$-wave scattering length $a_{\uparrow\downarrow}$. Once the dimer-dimer interaction becomes attractive for $g/|g_{\uparrow\downarrow}|\leq 2.2$, a many-body bound state gets formed\cite{PricoupenkoPetrov2018}, creating a droplet. This implies an important advantage of low-dimensional geometry, as experimentally there is no minimal number of atoms necessary for the creation of quantum droplets.

Another relevant physical question is how the size of a single dimer, $a_{\uparrow\downarrow}$, composed of two atoms belonging to the different components, compares to the mean interparticle distance. Apart from numerical factors $\sim 1$, the applicability condition for the perturbative theory assumes that $n_{0}a_{\uparrow\downarrow}\propto (\delta g/g)^{-2}\gg 1$, i.e., the size of a single dimer is large compared to the typical distance to the next particle. This means that, within the region of applicability of the Bogoliubov theory, the dimers can never be considered as separate single objects, and they only contribute to the collective properties. On the other hand, the existence of the droplets in the opposite limit of $\delta g/g\gg 1$ remains an open question, which should be addressed by non-perturbative approaches (such as Monte Carlo methods), or measured directly in some future experiment.

\subsection{The surface tension\label{Sec:surface tension}}

We conclude the study of stationary droplets by evaluation of the effective surface tension, a concept which is useful for understanding some of the dynamical properties which are considered below in Section~\ref{Sec:time dynamics}. For a 3D droplet, the surface tension can be extracted from the expansion of the energy density\cite{Beslic2009},
\begin{equation}
\frac{E}{N}=E_{v}+E_{s}N^{-1/3}+E_{c}N^{-2/3}\;,  \label{Eq:largeNdroplet}
\end{equation}
for $N^{-1}\rightarrow 0$. The coefficients in Eq.~(\ref{Eq:largeNdroplet}) define the volume, surface and curvature tension, respectively. The surface tension $\tau $ is related to $E_{s}$ as $\tau =E_{s}/(3\pi r_{0}^2)$, where the unit-volume radius $r_{0}$ is defined by condition $(4/3)\pi r_{0}^{3}=1$.

In the 1D system, the expansion parameter is $N^{-1}$, instead of $N^{-1/3}$ in Eq.~(\ref{Eq:largeNdroplet}), and the corresponding coefficients can be obtained analytically. The bulk energy density is $E_{v}=-2/9$, and the surface-energy coefficient is $E_{s}=16/(27e^2)$. In one dimension, the ``surface'' is reduced to two points, hence its size is independent of the size of the droplet. The respective surface tension is
\begin{equation}
\tau \equiv \frac{E_{s}}{2}=\frac{8}{27e^2} =0.040\;.
\label{Eq:surface tension}
\end{equation}

\section{Dynamics of the droplets \label{Sec:time dynamics}}

This section deals with collisions between droplets and excitation of a single one. In particular, as concerns the latter topic, we aim to produce the stability diagram in terms of the droplet's norm and wave number of the excitation.

\subsection{Collisions between two droplets\label{Sec:collisions}}

A soliton is nonuniform coherent wave self-trapped due to the action of the nonlinear dispersion. It is capable of maintaining its shape while moving at a constant velocity. In our case, the wave function, $\psi (x)$, given by Eq.~(\ref{exact:psi}) is shaped by the interplay of quadratic and cubic terms in Eq.~(\ref{Eq:GPE:full units}) for the droplet, similarly to the shape of the usual bright solitons. The time evolution of the wave function, $\psi(x,t)=\exp (-i\mu t)\psi (x,t)$, preserves the constant density profile, $n(x,t)=|\psi (x,t)|^2$, also in the case when the droplet is moving at a constant velocity.

Another distinct feature of a solitons in (nearly) integrable systems is that its shape must not be altered in a collision with another soliton. From this perspective, it is important to verify a possible persistence of the shape of droplets involved in the pairwise collisions. To address this issue, we simulated Eq.~(\ref{Eq:GPE}), using the split-step method based on the fast Fourier transform. The initial wave function was taken as a set of two counterpropagating droplets,
\begin{equation}
\psi (x,t\!=\!0)=e^{ikx/2+\varphi}\psi_{1}(x\!+\!x_{0})+e^{-ikx/2}\psi
_{2}(x\!-\!x_{0})
\end{equation}
where $\psi_{1}(x)$ and $\psi_{2}(x)$ are the stationary shapes of droplets with normalization $N_{1}$ and $N_{2}$, borrowed from Eq.~(\ref{exact:psi}), $\pm x_{0}$ are their initial positions, $\pm k/2$ are initial momenta of the colliding droplets, and $\varphi $ is the relative phase.

\begin{figure*}[tbp]
\begin{subfigure}{.49\textwidth}
  \centering
  \includegraphics[width=0.99\columnwidth,angle=0]{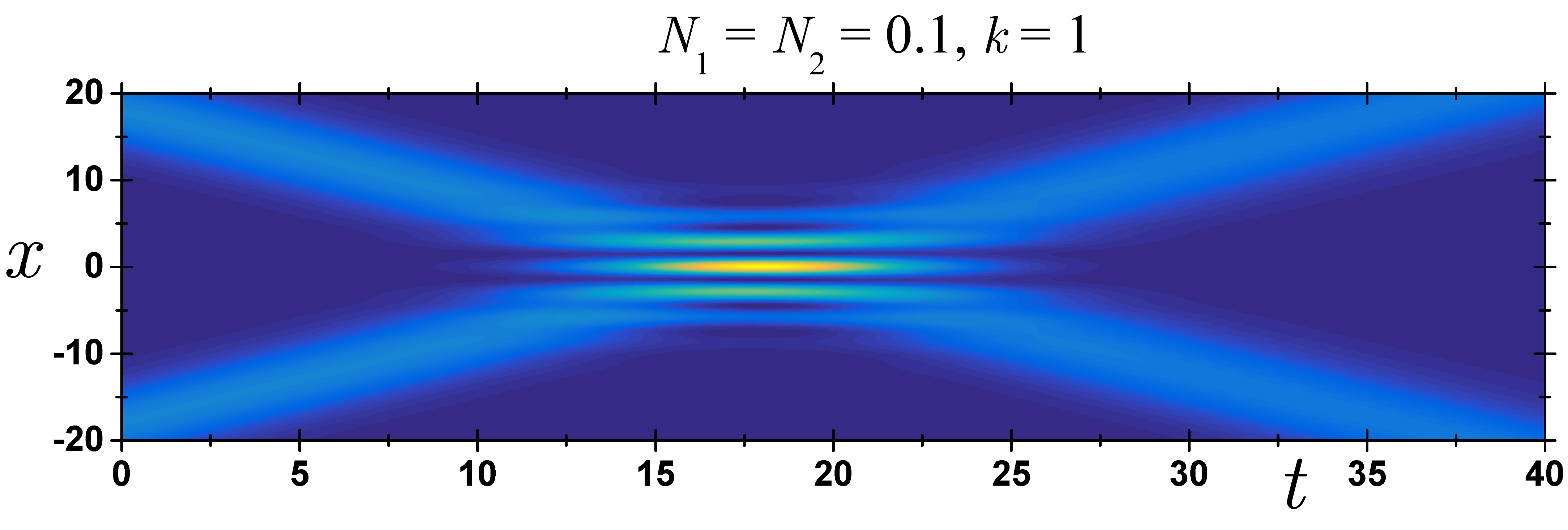}
  \caption*{(a)}
\end{subfigure}
\begin{subfigure}{.49\textwidth}
  \centering
  \includegraphics[width=0.99\columnwidth,angle=0]{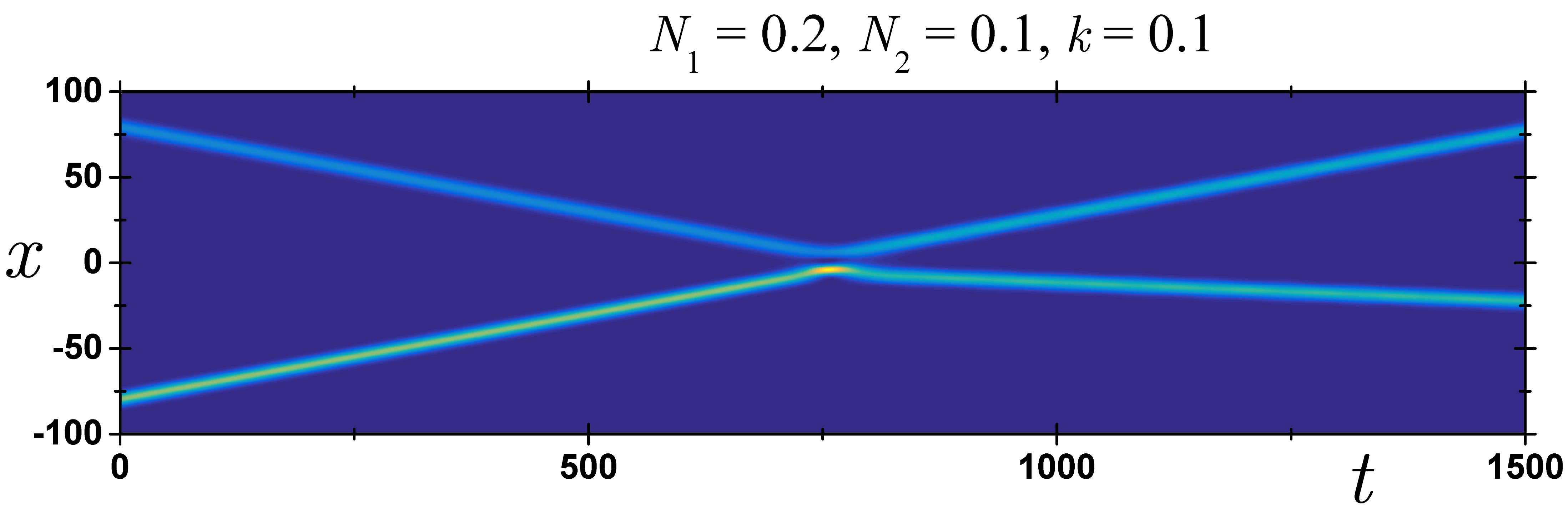}
  \caption*{(e)}
\end{subfigure}
\begin{subfigure}{.49\textwidth}
  \centering
  \includegraphics[width=0.99\columnwidth,angle=0]{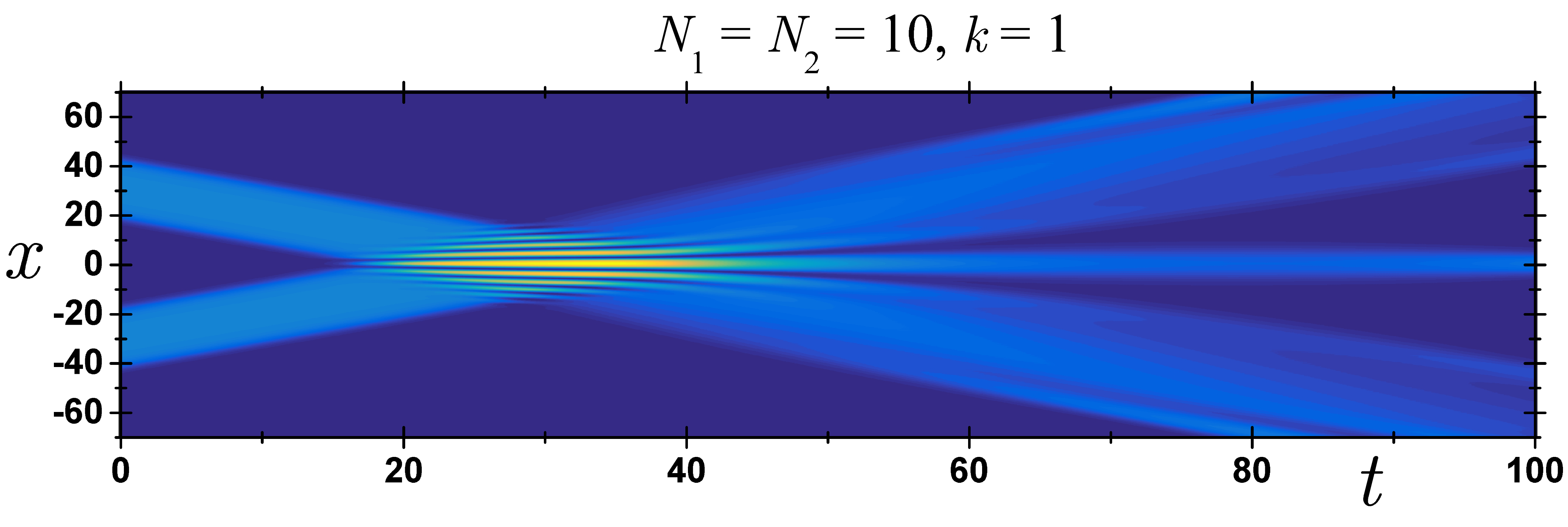}
  \caption*{(b)}
\end{subfigure}
\begin{subfigure}{.49\textwidth}
  \centering
  \includegraphics[width=0.99\columnwidth,angle=0]{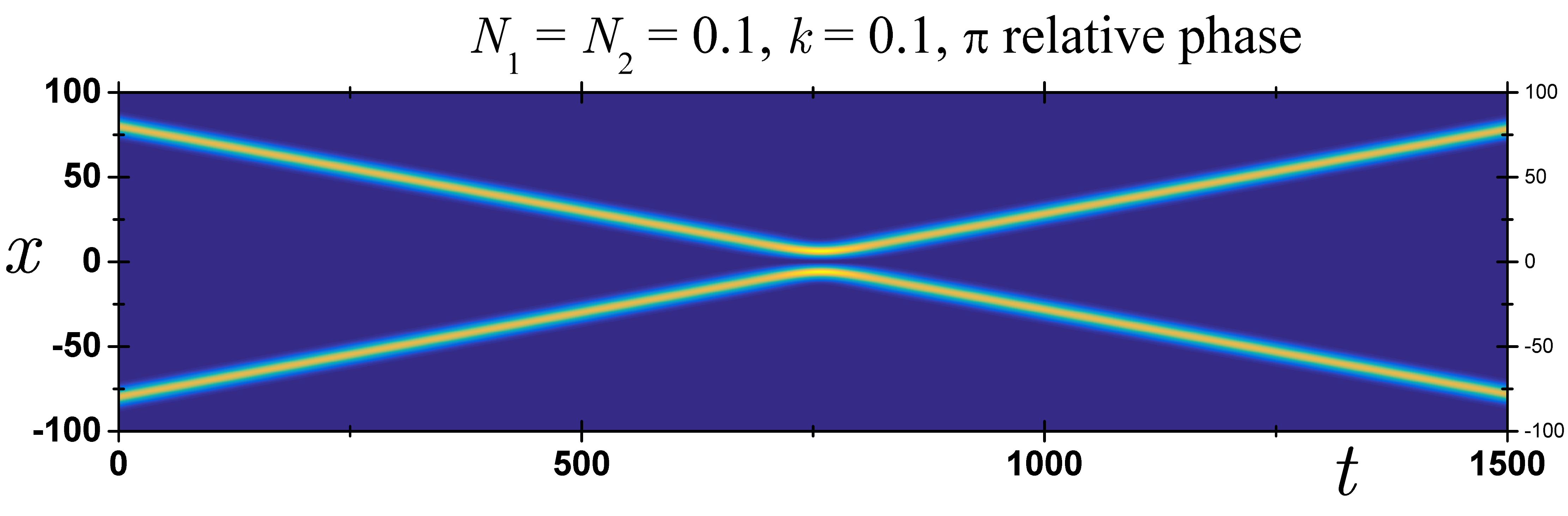}
  \caption*{(f)}
\end{subfigure}
\begin{subfigure}{.49\textwidth}
  \centering
  \includegraphics[width=0.99\columnwidth,angle=0]{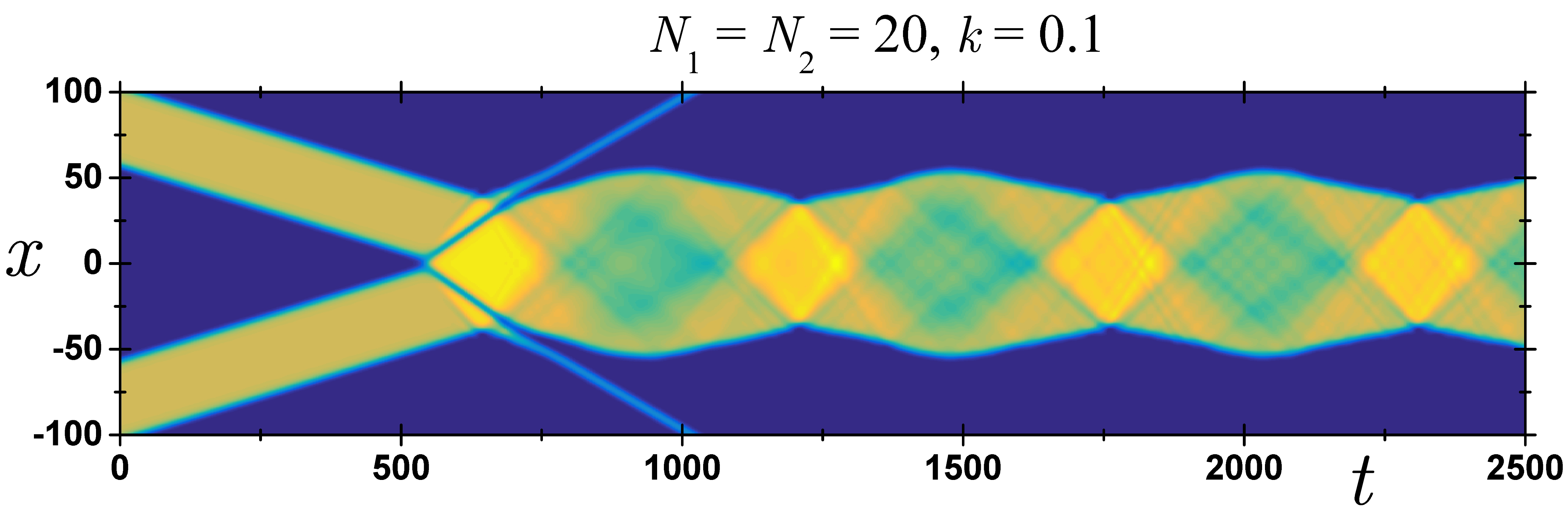}
  \caption*{(c)}
\end{subfigure}
\begin{subfigure}{.49\textwidth}
  \centering
  \includegraphics[width=0.99\columnwidth,angle=0]{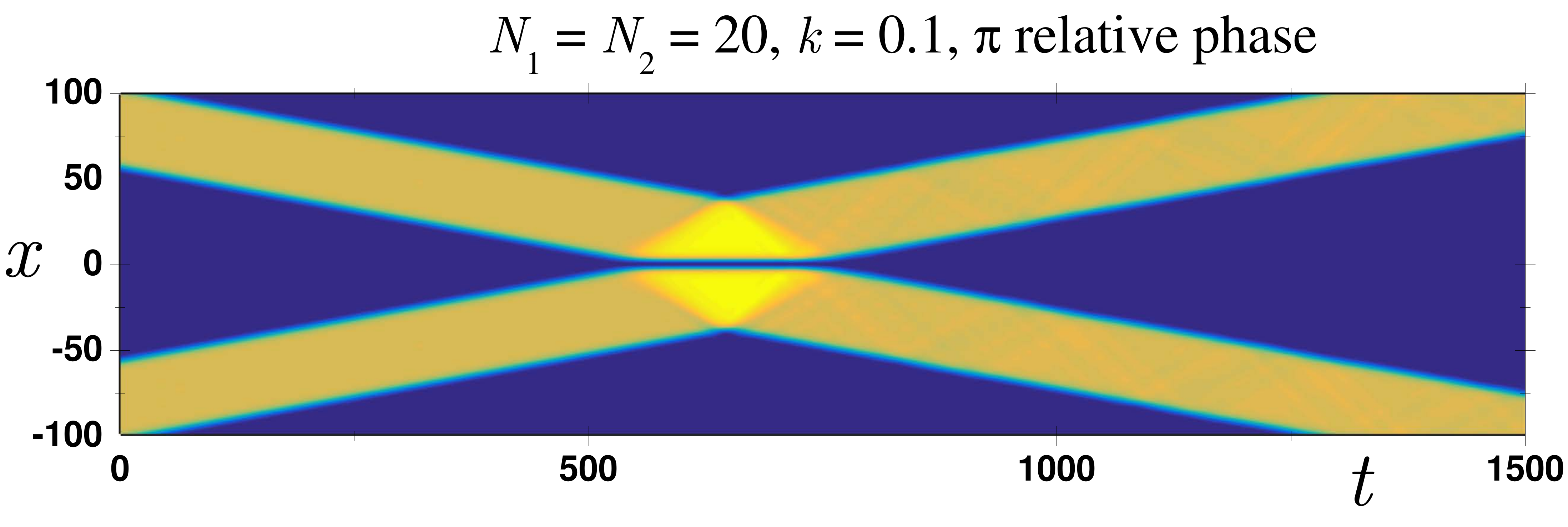}
  \caption*{(g)}
\end{subfigure}
\begin{subfigure}{.49\textwidth}
  \centering
  \includegraphics[width=0.99\columnwidth,angle=0]{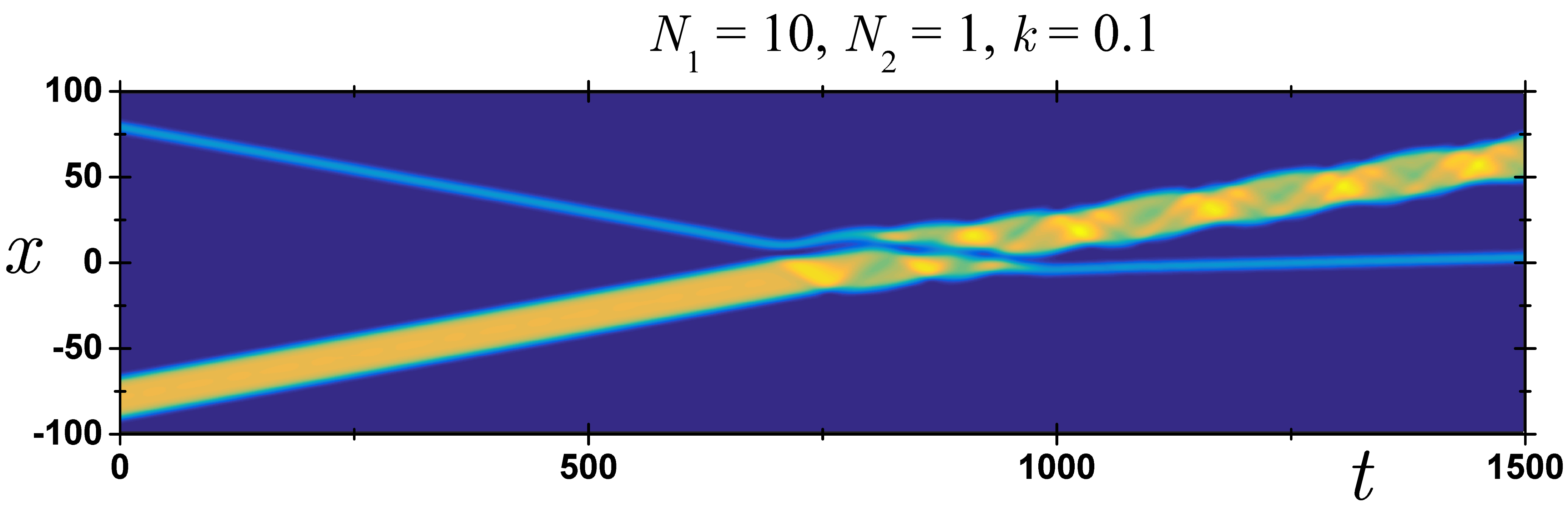}
  \caption*{(d)}
\end{subfigure}
\begin{subfigure}{.49\textwidth}
  \centering
  \includegraphics[width=0.99\columnwidth,angle=0]{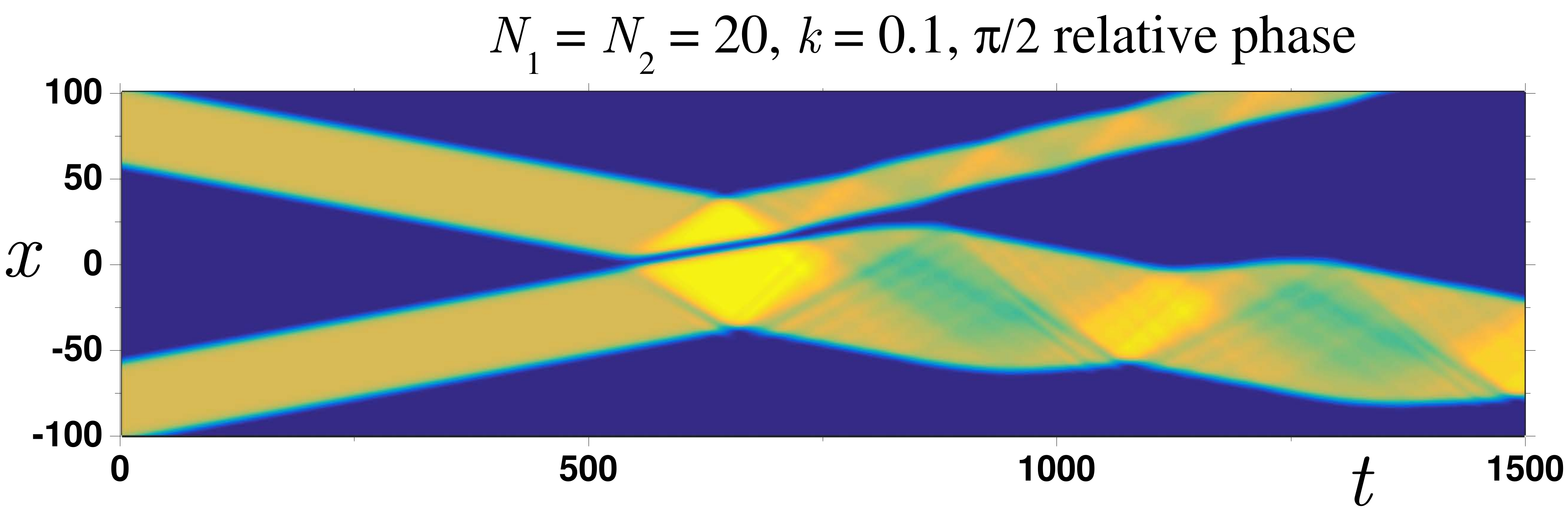}
  \caption*{(h)}
\end{subfigure}
\caption{(Color online) The density plot of the time evolution showing an interference pattern generated by collisions of two droplets, labeled by the normalizations ($N_{1}$ and $N_{2}$) and the incident momentum $k$.
All cases correspond to the in-phase collisions ($\varphi =0$), except when stated otherwise.}
\label{Fig:collision}
\end{figure*}

Figure~\ref{Fig:collision} shows density plots of the colliding droplets for a number of characteristic values of the relative momentum $k$ and norms $N_{1}$, $N_{2}$. The incident profile is the nearly Gaussian or a ``puddle'' one, for small and large $N$, respectively. In the former case, $N_{1}=N_{2}=0.1$, the shape of each droplet is precisely preserved after the collision. At the collision point, an interference pattern might appear [see Fig.~\ref{Fig:collision}(a)], being a distinctive feature of the interplay of coherent matter waves. The interference is well visible for $k\gtrsim 1/W$, where $W$ is the width of the droplet. As shown in Section~\ref{Sec:static properties}, the droplet's properties in this regime, $N\lesssim 1$, are well captured by the Gaussian-based VA.

The situation is quite different for large droplets, see Figs.~\ref{Fig:collision}(b,c). In this case, the shape of the droplet is no longer preserved after the collision. The example shown in Fig.~\ref{Fig:collision} (b) corresponds to the collision of fast-moving droplets ($k=1$, $N_{1}=N_{2}=10$), with the large momentum, in comparison to the droplet's inverse width. The interference pattern is clearly visible at the moment of the collision, resulting in the formation of \emph{three} outgoing droplets. Both incoming droplets undergo fragmentation, forming an additional quiescent one. The norm of the newly formed stationary droplet is small, with a majority of the particles being kept in the moving ones.

In the case of slowly moving droplets ($k=0.1$, $N_1=N_2=20$), shown in Fig.~\ref{Fig:collision}(c), a majority of the particles stay trapped in the newly formed central droplet, with only relatively small numbers of particles kept by the outgoing droplets. The merged central droplet is highly excited, showing large-amplitude oscillations. In the following subsection we address in more detail conditions under which a strongly excited droplet remains stable or suffers fragmentation.

We also analyze collisions of droplets with unequal norms in panels (d,e) of Fig.~\ref{Fig:collision}. In Fig.~\ref{Fig:collision}(d) we examine a scattering between two small Gaussian-like droplets. In this case no significant excitation is visible, the droplets rather behave as unperturbed objects. Still their scattering is not fully elastic, as the trajectories are affected by the collision. In Fig.~\ref{Fig:collision}(e) a large ``puddle'' droplet $N=10$ collides with a small droplet with $N=1$. It can be seen that the large droplet becomes highly excited, exhibiting internal periodic vibrations. On the opposite, the small droplet remains essentially in an unperturbed shape, although its trajectory is deflected.

We also study the effect of the relative phase on the collision. Figures~\ref{Fig:collision}(f,g) show example of out-of-phase scattering. The phase difference of $\pi$ is known to effectively induce repulsion between solitons\cite{Stegeman1999,Nguyen2014}. We see that both small (f) and large (g) droplets with equal norms indeed bounce back at the moment of the collision, so that the final trajectory can be interpreted as a total reflection.
While for phase $\pi$ there is no excitation of the equal-norm droplets, this is not the case for some other phase differences. Figure~\ref{Fig:collision}(h) shows two large droplets colliding with phase difference of $\pi/2$.
In this case the outcoming droplets can be excited and are unequally distributed in terms of the normalization.
The larger droplet is less deflected, similarly to the case (e) of unequal masses.
As the phase difference is decreased, the lateral outgoing droplet becomes smaller, see Figs.~\ref{Fig:collision}(g,h,c).
The breaking of the symmetry between the originally identical droplets in this case can be explained following the line of Ref.~\cite{KhaykovichMalomed2006}, as a consequence of the mismatch between the ``amplitude'' and ``phase'' centers of the two-droplet configuration with a phase shift different from 0 and $\pi$.
Eventually for zero difference, see case (c), the merged becomes stationary and two lateral droplets are formed.

\subsection{Intrinsic excitations in a single droplet\label{Sec:single droplet dynamics}}

In this subsection we study density-modulation excitation modes with a certain wave number in a single droplet, and address their stability.

\subsubsection{The breathing mode\label{Sec:breathing mode}}

The ``breathing'' or monopole mode is the lowest compression mode, with the spinor components moving in-phase, making the droplet size periodically oscillating. We excite this mode by driving a single droplet out of equilibrium and study the ensuing dynamics, simulating Eq.~(\ref{Eq:GPE}). The respective initial excitation is imposed by slightly changing the norm of stationary solution~(\ref{exact:psi}).

\begin{figure}[tbp]
\includegraphics[width=\columnwidth,angle=0]{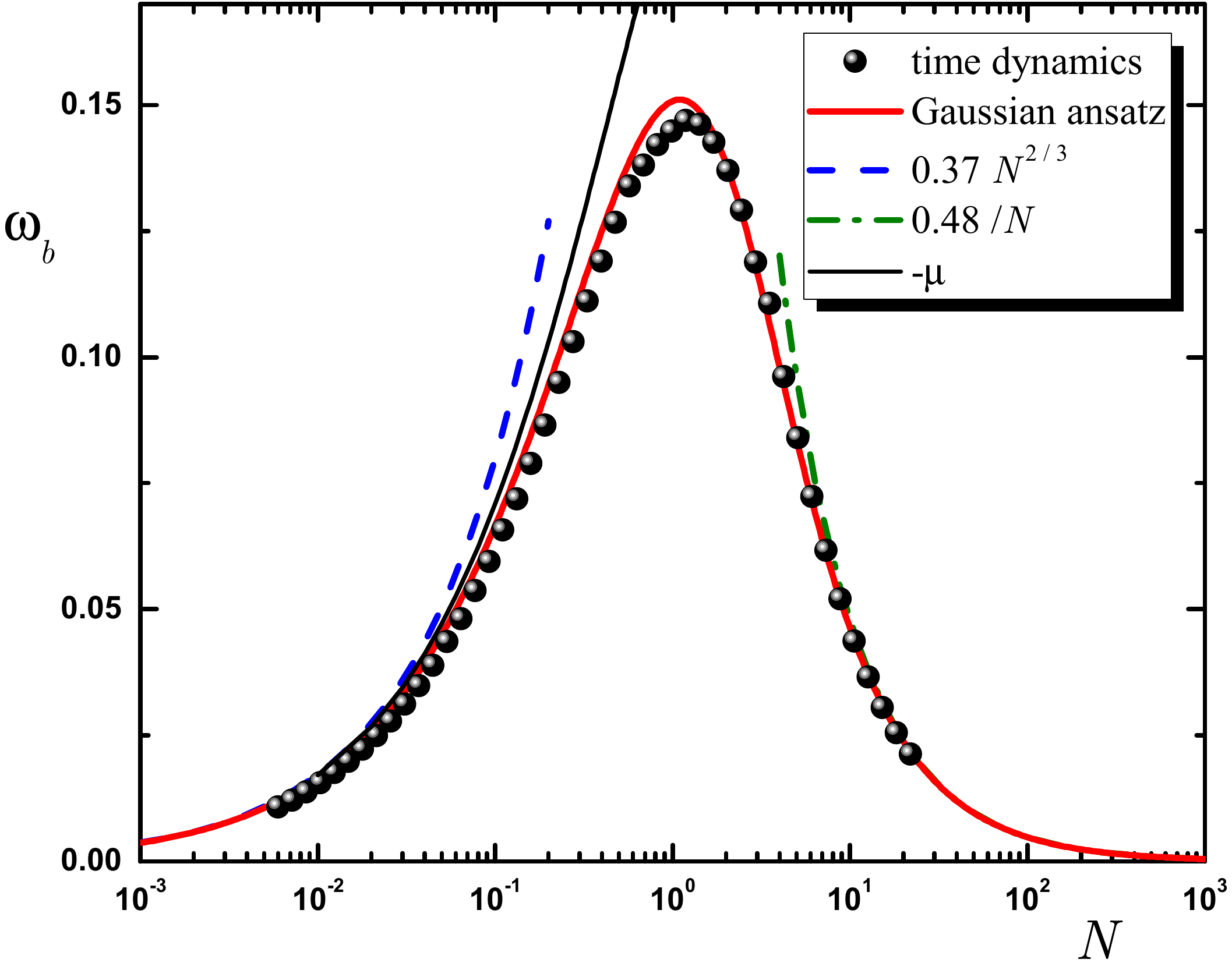}
\caption{(Color online) The frequency of the small-amplitude breathing mode excited in the single droplet, $\omega_{b}$, versus norm $N$. Symbols represent results of simulations of the Gross-Pitaevskii equation~(\ref{Eq:GPE}) with the input obtained by multiplying the norm of stationary solution~(\ref{exact:psi}) by perturbation factor $1.001$. The solid line depicts the prediction produced by the VA, as per Eq.~(\ref{Eq:freq}). The dashed and dashed-dotted lines correspond to the expansions for small and large $N$, as per Eqs.~(\ref{Eq:freq:small}) and (\ref{Eq:freq:large}), respectively.}
\label{Fig:breathing}
\end{figure}

Figure~\ref{Fig:breathing} presents the dependence of the resultant breathing mode on the droplet's size. There is a non-monotonous dependence with the largest stiffness reached around $N\approx 1$, when the droplet attains its minimal size. Symbols show results of a single-frequency fit to the density at the droplet's center. The solid and dashed lines show, respectively, the prediction produced by the VA, in the form of Eq.~(\ref{Eq:freq}), and asymptotic expansions obtained by means of the same method. While the underlying Gaussian \textit{ansatz} is expected to be quantitatively correct for $N\ll 1$, the overall agreement is remarkably accurate even for large ``puddle'' droplets, whose density profile has the flat-top shape.

In three-dimensional droplets for some parameters all excitation modes (both breathing and surface ones) have energy larger than the absolute value of the chemical potential\cite{Petrov2015}.
From energetic considerations for such parameters the droplet is not able to sustain any excitation and if an excited droplet is generated, it will lose atoms until all excitations are gone.
Such ``autocooling'' mechanism was argued to generate droplets in the true ground state, corresponding to the zero temperature\cite{Petrov2015}..
Instead, in one-dimensional geometry we find that $\hbar\omega_b < |\mu|$ for all parameters even if this condition is only marginally satisfied for small normalization $N\to 0$.
This implies that the ``autocooling'' mechanism is no longer applicable in one-dimensional geometry.

It is instructive to confront the breathing mode frequency in the puddle regime with the energy of a phonon with the minimal possible momentum.
The flat top can be excited to support linear phonons, $E(k) =  \hbar c |k|$, where $c$ is the speed of sound of the soft mode.
The minimal momentum can be approximated as $k \approx \pi / L$ where
$L = n_0 / N$ is the linear size of the droplet approximated here by the bulk density $n_0$.
This results in $\omega_b \propto 1/N$ scaling with the coefficient of proportionality of the order of one (at this level of accuracy the size $L$ is the same as the width $W$ of the Gaussian ansatz or the mean-square size $\langle x^2\rangle^{1/2}$).
The resulting functional dependence on $N$ is the same as for the breathing mode in the limit of $N\gg 1$, see Eq.~(\ref{Eq:freq:large}).
This finding is interesting as the ultradilute dipolar liquids were found to be essentially incompressible\cite{Pfau2016JPB} while in the present setting, the phonons can be excited on the flat top part of the puddle.

\subsubsection{The excitation initialized by density modulation with a finite wave number, and disintegration of the droplet\label{Sec:finite momentum}}

Another scenario of the excitation of internal dynamics of the droplet is provided by imprinting periodic density modulation onto it, with wave number $k$. The corresponding initial wave function is
\begin{equation}
\psi(x,t=0)=\psi_{\mathrm{exact}}(x)\cos(kx)\;,  \label{Eq:k}
\end{equation}
where $\psi_{\mathrm{exact}}(x)$ is the exact solution~(\ref{exact:psi}). As a result, in direct simulations the perturbed droplet may keep its shape entirety or suffer fragmentation, depending on $N$ and $k$.

\begin{figure}[tbp]
\includegraphics[width=\columnwidth,angle=0]{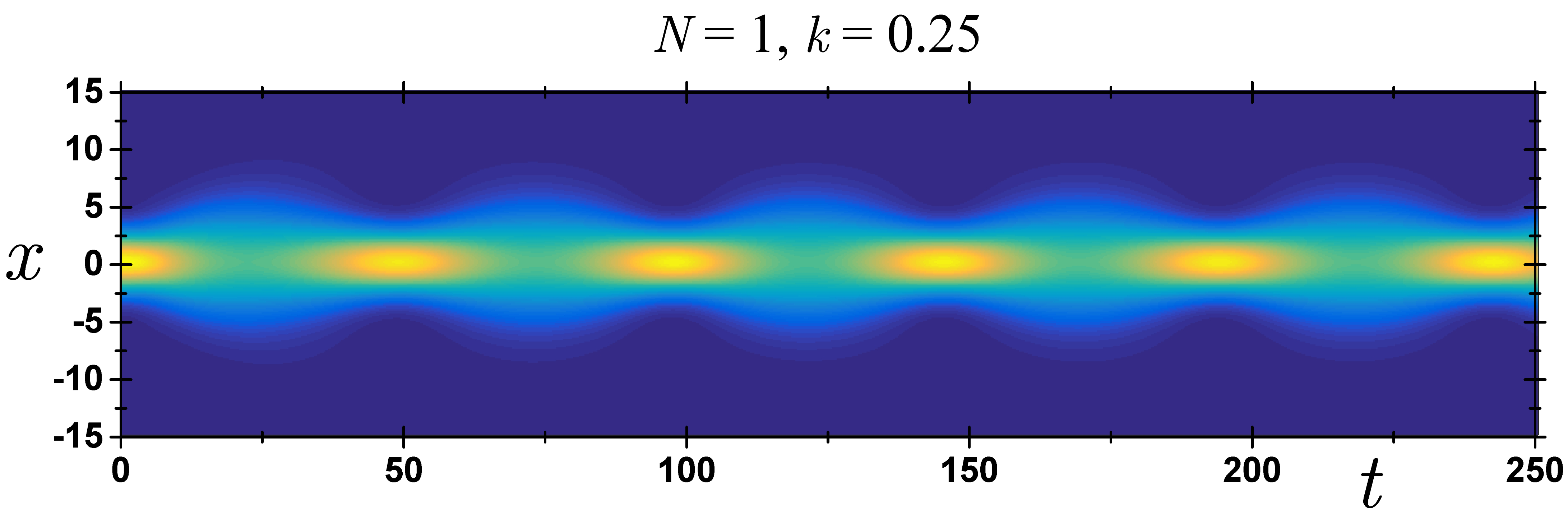}
\includegraphics[width=\columnwidth,angle=0]{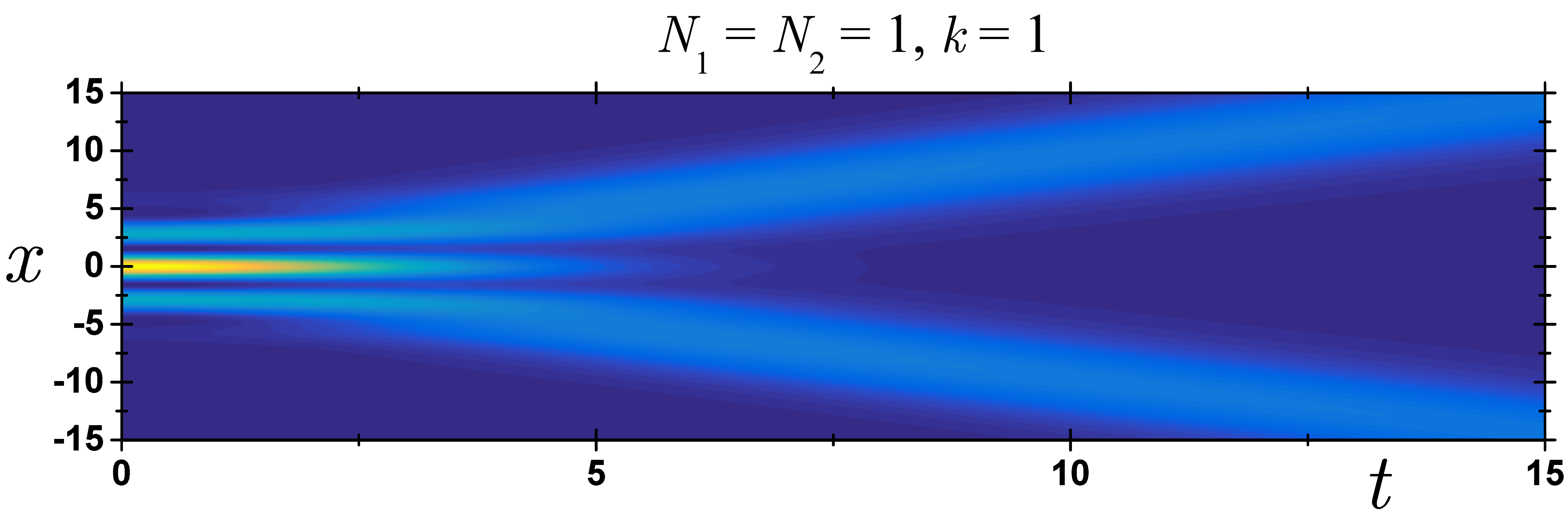}
\caption{(Color online) Density plots of the evolution of the droplet, initiated by the density modulation, imposed as per Eq.~(\ref{Eq:k}). Two characteristic examples (survival and fragmentation of the perturbed droplet) are shown. Both cases pertain to $N=1$, and $k=0.5$ in (a), or $k=1$ in (b).}
\label{Fig:excitation dynamics}
\end{figure}

Two possible outcomes of the evolution are illustrated by typical examples displayed in Fig.~\ref{Fig:excitation dynamics}. If the energy of the excitation is much smaller than the potential barrier induced by the surface tension, the droplet avoids fragmentation, as shown in Fig.~\ref{Fig:excitation dynamics}(a). In this case, almost periodic oscillations are observed in the width of the droplet. In the opposite limit of high excitation energy, the droplet splits in two or more escaping fragments (which are smaller droplets) which fly away, as shown in Fig.~\ref{Fig:excitation dynamics}(b).

\begin{figure}[tbp]
\includegraphics[width=\columnwidth,angle=0]{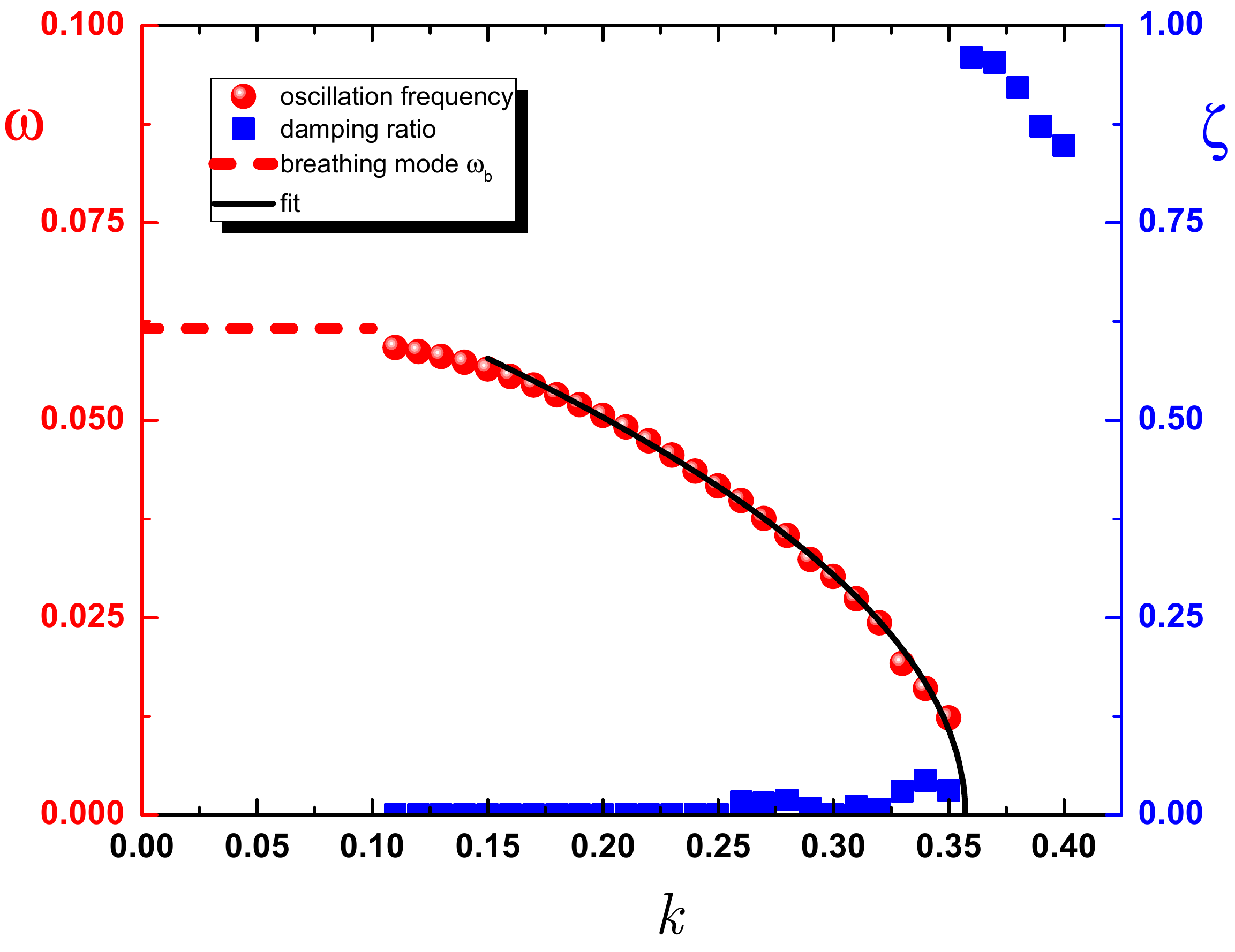}
\caption{(Color online) The frequency, $\omega $ (left axis), and damping ratio, $\zeta $ (right axis) of oscillations following the application of the density modulation to the droplet, as per Eq.~(\ref{Eq:k}), for $N=0.1$. The critical value of wave number, $k_{c}$, above which the perturbed droplet splits, is obtained by fitting the frequency in the stable region to $\omega (k)=a\sqrt{k_{c}-k}$. The fit is shown by the solid black line. The dashed line shows the frequency of the breathing mode, $\omega_{b}$, from Fig.~\ref{Fig:breathing}.}
\label{Fig:freq}
\end{figure}

To better understand the mechanism leading to the possible disintegration of a droplet, we study its oscillations following the application of the density-modulation momentum, see Fig.~\ref{Fig:excitation dynamics}(a), and measure their frequency and amplitude. To do so, we consider the density at the center of the droplet as a function of time and fit it to damped harmonic oscillations, $|\psi (x=0,t)|^2=A\cos (\sqrt{1-\zeta^2}\omega t)\exp (-\zeta \omega t)+B$, where $A$ is an amplitude, $B$ an offset, $f$ the frequency, and $0<\zeta <1$ the damping ratio. Typical dependencies of $\omega $ and $\zeta$ on $k$ are shown in Fig.~\ref{Fig:freq}. For small wave numbers $k$, the damping is absent, $\zeta \approx 0$, and the droplet oscillates with the frequency of a small-amplitude breathing mode $\omega_{b}$, as shown in Fig.~\ref{Fig:breathing}. For larger $k$, the frequency starts softening, and for $k$ exceeding a critical one, $k_{c}$, the droplet splits into several fragments. In this case, the density oscillations at the center exhibit strong damping, with the damping ratio approaching its largest value, $\zeta \approx 1$. In the stability region, $k<k_{c}$, the oscillation amplitude increases with $k$ and becomes so large that the density at the center may even vanish, thus making a hole in the condensate, which periodically opens and closes, while the droplet does not fall apart yet. The oscillation frequency vanishes in the limit of $k\rightarrow k_{c}$, which corresponds to the infinite period, so that, once the droplet splits in two fragments, they never recombine.

\subsubsection{The stability diagram\label{Sec:stability diagram}}

\begin{figure}[tbp]
\caption{(Color online) The stability diagram for a single droplet in the plane of norm $N$ and wave number $k$ of the initially applied density modulation~(\ref{Eq:k}). Symbols show the stability border $k_{c}$ obtained by fitting the oscillation frequency, cf. Fig.~\ref{Fig:freq}. Lines correspond to different values of the Weber number, defined as per Eq.~(\ref{W}): $\mathrm{We}=1$, $2$, and $3$ (dashed, dashed-dotted, and dashed-dotted-dotted lines, respectively).}
\includegraphics[width=\columnwidth,angle=0]{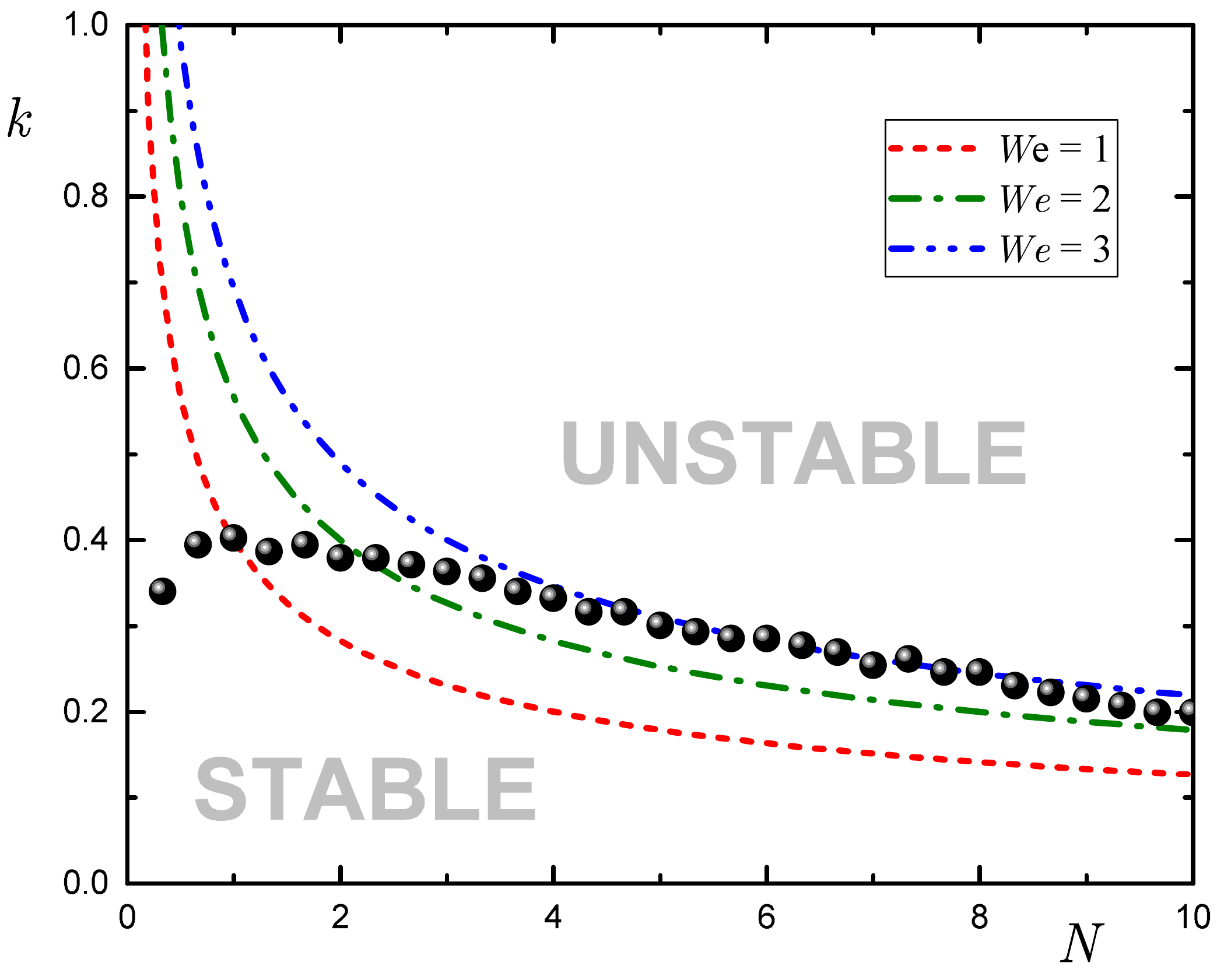}
\label{Fig:DiagStable}
\end{figure}

The stability diagram for the excited droplet in the plane of $\left(N,k\right) $ is displayed in Fig.~\ref{Fig:DiagStable}, in which symbols correspond to values $k_{c}$ extracted from systematic simulations of Eq.~(\ref{Eq:GPE}), according to the procedure outlined in subsubsection~\ref{Sec:finite momentum}. The droplet remains undivided at $k<k_{c}$. It is seen that the strongest stability corresponds to $N\approx 1$. The stability-threshold line may be interpreted in terms of energy considerations, by comparing the collisional kinetic energy\cite{Orme1997} associated with the imposed wave number, $E_{\mathrm{kin}}=N\hbar^2k^2/\left( 2m\right)$, and the surface energy, $E_{s}$, see Eq.~(\ref{Eq:surface tension}). The ratio of the two energies
\begin{equation}
\mathrm{We} = \frac{E_{\mathrm{kin}}}{E_{s}}=\frac{N\hbar^2k^2}{4m\tau}
\label{W}
\end{equation}
is known as the (modified) \textit{Weber number} \cite{Weber,Frohn2000}. Curved lines in Fig.~\ref{Fig:DiagStable} correspond to $\mathrm{We}=1,~2,~$and $3$. We find that, for $N\gtrsim 4$, the classical prediction based on a fixed value of the Weber number explains the stability diagram reasonably well. On the other hand, the stability for $N\lesssim 1$ is quite different. In this regime the perturbation with wave number $k$ may efficiently create an excitation in the droplet only at $k\gg 1/W$, where $W$ is the width of the droplet.

\section{Conclusion\label{Sec:conclusions}}

The main results reported in this paper are summarized as follows. We have studied static and dynamical properties of 1D two-component Bose gases forming quantum droplets, in the framework of the mean-field theory (GPEs) amended by the BMF (beyond-mean-field) corrections. Properties of the droplets greatly differ, depending on their norm $N$. Small droplets with $N\ll 1$ have an approximately Gaussian shape, being well described by the corresponding VA (variational approximation). Collisions between small droplets do not essentially alter their shape, hence droplets may be considered as solitons in a nearly integrable setting. On the other hand, large ``puddle'' droplets with $N\gg 1$ feature a top-flat density profile, with an approximately constant density corresponding to its equilibrium value in the uniform liquid. Although the VA fails to describe the exponential decay of the density profile at large distances, it is quite precise for small droplets and even produces meaningful results for a number of quantities of the ``puddle'' droplets. We have observed splitting and merger in collision of such extended droplets, depending on the collision velocity. We have produced the stability diagram for a single droplet with respect to imprinting a spatially periodic density modulation onto it. It demonstrates a fragmentation threshold in large (broad) droplets, with the critical Weber number $\sim 1$.

As an extension of the present work, it may be interesting to verify the validity of the mean-field theory, amended by the BMF terms, for predicting energies, density profiles and frequencies of oscillations, by means of the quantum Monte Carlo technique. In particular, it will be relevant to check if the entrainment between two superfluid components, known as the Andreev-Bashkin effect\cite{NespoloAstrakharchikRecati2017,ParisiAstrakharchikGiorgini2018}, can be observed in intrinsic oscillations of the droplets.

\section*{Acknowledgements}

We thank J. Boronat, D. S. Petrov and L. P. Pitaevskii for fruitful discussions. G.E.A. acknowledges partial financial support from the MICINN (Spain) Grant No.~FIS2014-56257-C2-1-P and FIS2017-84114-C2-1-P. The work of B. A. M. on this project was supported in part by the joint program in physics between NSF and Binational (US-Israel) Science Foundation through project No. 2015616, and by the Israel Science Foundation through Grant No. 1286/17. We thankfully acknowledge access to computer resources at MareNostrum and technical support provided by Barcelona Supercomputing Center (FI-2017-1-0009).

\bibliographystyle{apsrev4-1} 


%

\end{document}